\documentclass{aa}

\usepackage[dvips]{graphicx}

\newcommand{\gta}{\ga}
\newcommand{\lta}{\la}

\newcommand{\lprop}{\;
  \raise0.3ex\hbox{$\propto$\kern-0.75em\raise-1.1ex\hbox{$\sim$
  }}\;\hskip-2pt }

\newcommand\sfrac[2]{{\textstyle{\frac{#1}{#2}}}}

\newcommand{\kms}{\,{\rm km\,s^{-1}}}
\newcommand{\cm}{\,{\rm cm}}
\newcommand{\cmcube}{\,{\rm cm^{-3}}}
\newcommand{\G}{\,{\rm G}}
\newcommand{\gcmcube}{\,{\rm g}\,{\rm cm^{-3}}}
\newcommand{\mG}{\,{\rm mG}}
\newcommand{\mkG}{\,\mu{\rm G}}

\newcommand{\kpc}{\,{\rm kpc}}
\newcommand{\ppc}{\,{\rm pc}}
\newcommand{\yr}{\,{\rm yr}}
\newcommand{\s}{\,{\rm s}}

\begin{document}

\title{Galactic dynamos with captured magnetic flux and an accretion flow}

\author{David Moss$^1$
\and
Anvar Shukurov$^2$}

\offprints{D.~Moss}

\institute{
Department of Mathematics, University of Manchester, Manchester M13~9PL,
U.K.
\and
Department of Mathematics, University of Newcastle, Newcastle
  upon Tyne NE1~7RU, U.K.\\
}

\date{Received ...; accepted ...}

\thesaurus{04(02.13.1; 02.13.2; 09.13.1; 10.03.1; 11.09.4; 11.13.2;
11.19.2)}

\authorrunning{D.~Moss and A.~Shukurov}

\titlerunning{Galactic dynamos with captured magnetic flux}

\maketitle

\begin{abstract}
We examine the behaviour of an axisymmetric galactic dynamo model
with a radial accretion flow in the disc. We also introduce a
vertical magnetic flux through the galactic midplane, to simulate the
presence of a large scale magnetic field trapped by the galaxy when
forming.  The trapped vertical flux is conserved and advected towards
the disc centre by the radial flow.  We confirm that accretion flows
of magnitude  several km\,s$^{-1}$ through a significant part of the
galactic disc can markedly inhibit dynamo action.  Moreover,
advection of the vertical flux
in general results in
mixed parity galactic fields. However, the effect is nonlinear and
non-additive -- global magnetic field energies are usually
significantly smaller that the sum of purely dynamo generated and
purely advected field energies.  For large inflow speeds, a form of
`semi-dynamo' action may occur.

We apply our results to the accumulation and redistribution, by a
radial inflow, of a vertical magnetic flux captured by the Galactic
disc.  Taking representative values, it appears difficult to obtain
mean vertical fields near the centre of the Milky Way that are much
in excess of $10\mkG$,
largely because the galactic dynamo and
turbulent magnetic diffusion modify the external magnetic field
before it can reach the disc centre.

\end{abstract}

\section{Introduction}                  \label{intro}
Dynamo systems only rarely occur isolated magnetically from their
environment.  To assume such isolation may be a plausible assumption
for, e.g., stellar dynamos, but the situation is not so clearcut for
spiral galaxies.  Nevertheless, it is generally supposed that the
dynamo acts independently of any trapped vertical flux, and the
interaction of a dynamo system with its magnetic environment has
rarely been addressed. However, the effects of external magnetic
fields have been invoked to explain the origins of various features
of astrophysical discs, such as vertical dust filaments in spiral
galaxies (Sofue 1987), strong vertical magnetic fields in the
centre of the Milky Way (Sofue \& Fujimoto 1987, Chandran, Cowley \&
Morris 2000) and, in a different astrophysical (but not conceptual)
context, the launching and collimation of outflows in accretion discs
around young stars and compact objects (Blandford \& Payne 1982,
Mestel 1999). It is widely accepted, however, that both accretion
discs and the discs of spiral galaxies are sites of dynamo action
(Parker 1979; Zeldovich, Ruzmaikin \& Sokoloff 1983; Ruzmaikin,
Shukurov \& Sokoloff 1988; Beck et al.\ 1996; Kulsrud 1999; Mestel
1999).  In this paper we discuss, in a galactic context, the
nonlinear interaction between a disc dynamo and a trapped vertical
flux redistributed by the radial inflow of matter.

The most important aspect of the problem in the galactic context is
the origin of the vertical magnetic field in the Galactic centre
where it can be as strong as $1\,$mG within the inner $200\ppc$
radius, implying vertical magnetic flux of about $40\G\ppc^2$ if the
field is unidirectional and pervasive (Morris \& Serabyn 1996).  The
field geometry is apparently dipole-like, in contrast to the
quadrupole symmetry dominant near the Sun. With a gas density of
10--$100\cmcube$ between molecular clouds and the turbulent
velocities of 20--$30\kms$ (e.g., Morris \& Serabyn 1996), the field
is 100--10 times above the equipartition level with the turbulent
kinetic energy. These features suggest that  the vertical
magnetic field in the Galactic centre might arise from an accumulation
of an external magnetic flux advected from large radii over the
Galactic lifetime (Sofue \& Fujimoto 1987, Chandran et al.\ 2000).
However, the Galactic disc is magnetically active, being a site of
dynamo action, and so it is not clear whether or not any vertical
magnetic field can be dragged through it.  Our results suggest that
it is rather implausible that the vertical magnetic flux at the
Galactic centre results from the redistribution of magnetic field
trapped at an early stage of Galactic evolution.

\section{Formulation of the problem}
\label{formul}
The interaction of a nonlinear dynamo system with an external
magnetic field can be non-trivial.  Drobyshevski (1977) proposed  the
idea of the `semi-dynamo', a system which amplifies magnetic field at
the expense of kinetic energy like a self-excited dynamo, but in
which the field would decay without the presence of an external
magnetic field.
Convective dynamos in the presence of an imposed magnetic field have
been studied by Sarson et al.\ (1997) and Sarson, Jones \& Zhang
(1999) in application to magnetic fields of Io and Ganymede, Jovian
satellites orbiting within the giant planet's magnetosphere. These
authors report on magnetic states with self-maintained magnetic field
which, however, revert to a state of nonmagnetic convection if the
external field is withdrawn or weakened.  In this model, a uniform
magnetic field parallel to the rotation axis is included as a source
term in the induction equation, although the results remain the same
when the external magnetic field is introduced through boundary
conditions (G.~R.~Sarson, private communication). There is a
plausible connection between the behaviour of these systems and  that
of Drobyshevski's semi-dynamos.

There have been a number of studies of stellar dynamos with a fixed
magnetic field as boundary condition at the bottom of the convection
zone, simulating the presence of an interior field frozen in to the
radiative core (e.g.\  Pudovkin \& Benevolenskaya 1984, Boyer \& Levy
1984, Moss 1996).  The effect here was less fundamental, being to
change properties such as parity, rather than cause semi-dynamo
action.

A two-dimensional mean-field dynamo with external magnetic field has
been considered by Reyes-Ruiz \& Stepinski (1997) who, however,
neglected any radial velocity and obtained a solution of the
nonlinear ($\alpha$-quenched) dynamo equations with external magnetic
field by merely adding a constant vertical field to the solution of
the standard dynamo equations. The physical importance of their
results for realistic degrees of nonlinearity is therefore
questionable.

The inward advection (dragging) of an external magnetic field by
accretion flow in a Keplerian disc has been considered by Lubow,
Papaloizou \& Pringle (1994) (numerical solutions have been presented
by Reyes-Ruiz \& Stepinski 1996). These authors disregarded any
dynamo action and considered only the balance between advection of a
purely poloidal magnetic field and radial diffusion.  Their results
show that efficient advection of an external magnetic field into a
viscous disc is only possible if the magnetic diffusivity is much smaller
than the kinetic. Otherwise inward advection is balanced by radial
magnetic diffusion and the steady-state magnetic field in the disc
can only be weak.

There are several possible types of external magnetic field for
galaxies and accretion discs. This can be a field of cosmological
origin. Current models of magnetic field generation in the early
Universe predict that the strength of a primeval field at the
galactic scales can only be negligible (e.g.\ Beck et al.\ 1996), so
this option has lost its importance. For galaxies belonging to a
galaxy cluster (but not the Milky Way), the intracluster magnetic
field (Kronberg 1994) can be a significant source of external
magnetic flux. The strength of this field is about $1\mkG$, but only
a fraction of it may have the appropriate poloidal structure in the
galactic frame.  Away from rich galaxy clusters, the observational
upper limit on the cosmological magnetic field
($<3\times10^{-12}$\,G, Sciama 1994, Beck et al.\ 1996) implies an
upper limit of $3\times10^{-10}\,$G for a quasi-uniform magnetic
field captured by and compressed in a collapsing protogalaxy (for
radial compression by a factor 10).  For accretion discs in an active
galactic nucleus or in a stellar system, the interstellar magnetic
field of the parent galaxy ($\simeq1\mkG$ for the total field of
which a fraction would be poloidal field as seen by the disc) can
serve as an external field.  Stronger external magnetic fields
originating in the central star can occur in accretion discs of
binary stellar systems (e.g.\  Bardou \& Heyvaerts 1996).

Although the relative orientation of the disc plane and the external
magnetic field can be arbitrary, we neglect any magnetic field
component parallel to the disc plane and  allow only for a component
parallel to the rotation axis. Any horizontal component will be twisted by
differential rotation until the separation of magnetic line segments
of alternating sign becomes small enough for the field to be
destroyed by turbulent magnetic diffusion and/or reconnection
(e.g. Sects.\ 3.8 and 3.9 of Moffatt 1978).  The
only way to avoid or postpone the expulsion is by reducing the
magnetic diffusion as, e.g., in Howard \& Kulsrud (1997) and Chandran
et al.\ (2000), who only consider the effects of a relatively weak
ambipolar diffusivity.  Observations directly indicate the
presence of intense
turbulent motions in the interstellar gas of spiral galaxies (Minter
\& Spangler 1996). Even though interstellar magnetic fields are
dynamically important, there are no apparent indications of any
modification of the turbulence that could result in any significant
suppression of turbulent diffusion.  Furthermore, Howard \& Kulsrud
(1997) neglect magnetic buoyancy, which can be efficient in removing
horizontal magnetic field from the disc on a time scale of order the
Alfv\'en crossing time over the disc scale height.  Magnetic buoyancy
possibly  presents insurmountable difficulties for a horizontal
primordial magnetic field, whereas galactic dynamo action may even be
enhanced by buoyancy (Parker 1992; Moss, Shukurov \& Sokoloff 1999);
however we neglect the effects of magnetic buoyancy here so as to
keep the model as simple as possible.

Thus, any horizontal external magnetic field will be unimportant.
However, the behaviour of the vertical magnetic field $B_z$ is
different.  For a mean velocity field (including rotation) that is
independent of the vertical coordinate $z$, a uniform $B_z$ is not
distorted by the flow and is not subject to vertical diffusion (since
$\partial^2 B_z/\partial z^2\equiv0$).  Therefore, such a field can
be dragged into the disc by a radial flow. The dragging is especially
efficient if the disc is surrounded by vacuum (where magnetic
diffusivity is effectively  infinite).  The pile-up of the vertical
magnetic field will be opposed and balanced by the radial magnetic
diffusion.  The resulting steady state was discussed by Lubow et al.\
(1994). A $z$-dependent rotation would add an axisymmetric azimuthal
magnetic field produced from $B_z$ by the vertical shear.  In
addition, an azimuthal field can be produced by differential rotation
from a radial field generated by a vertical gradient in the inflow
velocity.

This picture is very much enriched by any dynamo action in the disc.
The dynamo generates its own magnetic field that can interact with
the intruding vertical flux either constructively or destructively.
The basic dynamo mode in a galactic disc (where the $\alpha$-effect
is positive in the northern hemisphere) is of quadrupolar parity
(see, e.g., Parker 1979, Zeldovich et al.\ 1983, Ruzmaikin et al.\
1988), opposite to the parity of the (quasi-)uniform vertical
magnetic field.  Therefore, the two fields can only interact via
nonlinear effects.  The situation can be different in accretion discs
where magneto-rotational instability can produce a negative
$\alpha$-effect resulting in a predominantly dipolar dynamo field
(Brandenburg, Nordlund \& Stein 1995).  The synergy of the vertical
flux and the dynamo may result in physically nontrivial effects such
as the semi-dynamo of Drobyshevski.
Some of the behaviours obeserved in our models for large
inflow speed can be considered as a nonlinear analogue of semi-dynamo
action wherein a subcritical dynamo action modifies non-trivially an
invading external magnetic field.

Radial flows can affect the disc dynamo in two obvious ways, both
suppressive. As discussed by Moss, Shukurov \& Sokoloff (2000), the
radial advection hinders dynamo action, basically because the
advection distorts the field from the eigenfunction.
This effect can be important when the radial velocity
$u_r\ga2\eta/h\simeq1\kms$, for parameter values typical of the Solar
neighbourhood ($h\simeq500\ppc$ is the scale height of the ionized
layer, $\eta\simeq10^{26}\cm^2\s^{-1}$ is the turbulent magnetic diffusivity).
The second effect is due to the contribution of an imposed
magnetic flux, dragged by the flow, to the nonlinear saturation of
the dynamo action, customarily modelled by the $\alpha$-quenching.
Now the strength of the $\alpha$-effect, responsible for the dynamo
action, is reduced by the magnetic field as
$\alpha\propto(1+B^2/B_0^2)^{-1}$, where $B$ is the strength of the
local mean magnetic field, comprising both the dynamo and advected
imposed flux. Here $B_0$ is a reference field strength at which
nonlinear dynamo effects become pronounced.  The suppression of the
dynamo action by the intruding magnetic field can be expected to be
significant if it becomes comparable to $B_0$.

All these effects cannot easily be separated, and it is difficult to
assess their possible importance without the detailed analysis attempted
here. In particular, our results presented below show that the dynamo
can sometimes survive a relatively strong imposed vertical flux --
the dynamo can fight against its suppression.

As discussed by, e.g.,  Lacey \& Fall (1985) and Moss et al.\ (2000),
the main mechanisms resulting in angular momentum transport with
ensuing systematic radial flow in spiral galaxies are (in order of
importance) the nonaxisymmetric gravitational field of a bar or
spiral arms (Lubow, Balbus \& Cowie 1986), viscous and magnetic
torques (Mestel 1999), and the infall of matter onto the disc (Pitts
\& Tayler 1996). The magnitude of the radial velocity and its radial
profile are uncertain; the plausible values range from 1 to $10\kms$
at a distance of order 3--10\,kpc from the centre in a spiral galaxy
such as the Milky Way.

We examine two basic galaxy models. One has a gas density
distribution taken from Milky Way observations in $r>2$ kpc (see
Sect.~\ref{galmod}) but, in order to allow examination of a wider
range of parameters we limited the central gas density in $r\leq 2$
kpc to unrealistically small values. These calculations are described
in Sects.~\ref{standard} and \ref{altmodel}.  In Sect.~\ref{SM} we
take a more realistic value of the central gas density, which is very
strongly peaked in $r<2$ kpc, and present results for a limited range
of parameters applicable to the Milky Way.  Thus readers who are only
interested in applications to the Milky Way might consider omitting
reading Sects.~\ref{standard} and \ref{altmodel}.

We emphasize that we consider a prescribed dipolar flux, trapped by
the disc at the time of galaxy formation. In reality, a galaxy is
likely to trap a mixed parity flux, but the quadrupolar component
will merely act as a seed for the dynamo.

\section{The model} \label{model}

\subsection{Basic equations}

We study a dynamo active disc embedded in a passive, conducting halo,
and solve the standard $\alpha^2\omega$ dynamo equation in the disc,
\begin{equation}
\frac{\partial \vec{B}}{\partial t}
        =\nabla\times(\vec{v}\times\vec{B}+\alpha\vec{B})
        -\nabla\times(\eta\nabla\times\vec{B})\;,
\label{mfd}
\end{equation}
but introduce an inwardly directed axisymmetric radial accretion flow
and an axisymmetric vertical  magnetic flux. The $\alpha$-effect and
turbulent magnetic diffusion are specified via scalar fields $\alpha$
and $\eta$. We study axisymmetric solutions of Eq.~(\ref{mfd}) in the
domain $-Z\leq z\leq Z$, $r_{\rm min}\leq r\leq R$, and define the
aspect ratio of the domain as
$\xi=Z/R$;,
where $r$ and $z$ are cylindrical polar coordinates.  We choose
\begin{eqnarray*} \label{alpha}
\alpha=
\cases{\displaystyle\frac{\alpha_0\sin(\pi z/h)}{1+\vec{B}^2/4\pi k\rho
v_{\rm t}^2}\;,&$|z|\leq h\;,$\cr
0\;,&$|z|>h\;,$\cr}
\end{eqnarray*}
defining the semi-thickness of the dynamo active disc, $h$, assumed
to be independent of $r$ (of course, $h<Z$).  Here $\alpha_0$ is the
background (unquenched) magnitude of the $\alpha$-effect, the gas
density is denoted $\rho$ (a function of $\vec{r}$), and $v_{\rm t}$
is the turbulent velocity.  We also define the disc aspect ratio
$\lambda=h/R$;.
We have thus introduced a conventional $\alpha$-quenching
nonlinearity, with $k$ a constant of order unity (reflecting some of
the uncertainty in our ideas about the reaction of the large scale
magnetic field on to the turbulent motions, but ignoring ongoing
controversies about the nature and effectiveness of the feedback of
the dynamo generated fields on the $\alpha$-effect -- see, e.g.,
Vainshtein \& Cattaneo 1992, Blackman \& Field 2000, Kleeorin et al.\
2000, Brandenburg \& Subramanian 2000).

Since we will consider only axisymmetric solutions, we can write
\begin{equation}
\vec{B}=B_\phi\widehat{\vec{\phi}}
        +\nabla\times(A_\phi\widehat{\vec{\phi}})\;,
\end{equation}
where $B_\phi$ and $A_\phi$ are the azimuthal components of the
magnetic field and vector potential, respectively. We also put
\begin{equation}
\vec{v}=u(\vec{r})\widehat{\vec{r}}+r\Omega(r)\widehat{\vec{\phi}}
        +\vec{u}_{\rm dia}\;,
\label{vels}
\end{equation}
where $\vec{u}_{\rm dia}=-\frac{1}{2}\nabla\eta$ allows for the
turbulent diamag\-net\-ism (Zeldovich 1956, Roberts \& Soward
1975), $u$ is the prescribed radial inflow, and $\Omega$ is the
galactic angular velocity.

For the magnetic diffusivity, we set
\[
\eta=
\cases{\eta_0\;,&$|z|\leq h\;,$\cr
\eta_0\left[1+(\eta_1-1)\Theta(z)\right]\;,&$|z|>h\;,$\cr}
\]
where $\eta_0$ is a constant,
\[
\Theta(z)=1
-\exp\left[-\frac{(|z|-h)^2}{l_\eta^2}\right]
\]
is a smoothed step function and, rather arbitrarily,
$l_\eta=\sfrac12(\xi-\lambda)$. Thus, the constant $\eta_1$ is the
ratio of the diffusion coefficient high in the halo to that in the
disc. We consider models with the turbulent magnetic diffusivity in
the halo significantly larger than in the disc.  This appears to be a
reasonable assumption for a galactic disc surrounded by a hot,
turbulent halo (Sokoloff \& Shukurov 1990; Poezd, Shukurov \&
Sokoloff 1993) as well as for an accretion disc corona supported by
turbulent pressure (Ouyed \& Pudritz 1997).  Anyway, the larger is
the magnetic diffusivity outside the disc, the easier is the dragging
of the vertical magnetic flux.

\begin{figure}
\centerline{\includegraphics[width=8.5cm]{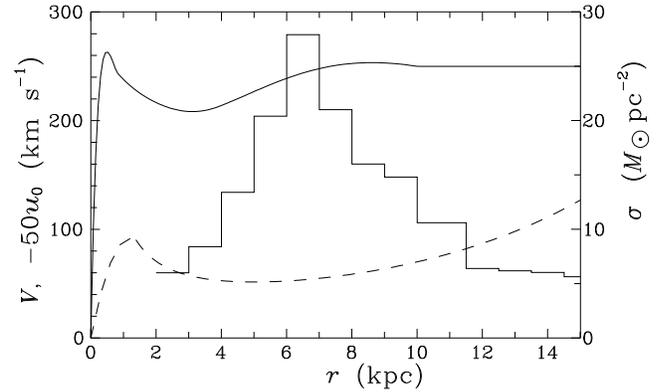}}
\caption[]{\label{Omeg}
The radial dependence of the Galactic rotation velocity $V=r\Omega$
(solid) (Burton \& Gordon 1978), the radial velocity in the disc
midplane $u_0$ (multiplied by $-50$), Eq.~(\protect\ref{inflow})
(broken), and gas surface density $\sigma$ at $r\geq2\kpc$ (Sanders
et al.\ 1984) (histogram).
}
\end{figure}

\subsection{A model for the Galaxy}
\label{galmod}
We take $\Omega$ from a model of the Milky Way (Burton \& Gordon
1978; see also p.~123 of Ruzmaikin et al.\ 1988), and show
$V=r\Omega(r)$ in Fig.~\ref{Omeg}. The surface density of the
interstellar gas $\sigma$ is taken from Sanders, Solomon \& Scoville
(1984), and also shown in Fig.~\ref{Omeg} for $r\geq2\kpc$. According
to Sanders et al.\ (1984) (see also Dame 1992), the surface density
of molecular hydrogen increases to $64\,M_\odot\ppc^{-2}$ at
$r=1\kpc$ and a few hundred $M_\odot\ppc^{-2}$ inside that radius.
In our calculations we use both a central density of
$300\,M_\odot\ppc^{-2}$ and also a naive extrapolation of
Fig.~\ref{Omeg} to $5\,M_\odot\ppc^{-2}$ at $r=0$.  The latter value
is used to allow exploration of general properties of dynamos with
inflow without the necessity of using a very high radial resolution.
These distributions of $\rho$ and $r\Omega$ describe a rather
old-fashioned model of the Milky Way, with the Solar galactocentric
distance and rotation speed being $10\kpc$ and $250\kms$, rather than
the currently accepted $8.5\kpc$ and $220\kms$, but we do not believe
that this choice affects our conclusions.  We neglected any possible
$z$-dependence of $\Omega$ outside of the disc. This is rather
uncertain, and would not dramatically affect the computed fields,
given that the inflow and dynamo action are confined to the disc
region, and so also is the bulk of the magnetic field.  We neglect
the possibility of any dynamo action in the halo (Sokoloff \&
Shukurov 1990; Brandenburg et al.\ 1992, 1993).

With $\eta$ growing with $z$, turbulent magnetism, resulting in
effective magnetic field transport at a velocity $\vec{u}_{\rm dia}$,
compresses magnetic field to the disc (Brandenburg et al.\ 1992;
Gabov, Sokoloff \& Shukurov 1996). However, this effect is relatively
unimportant.

Moss et al.\ (2000) discuss a crude but plausible model for the
radial profile of a radial flow in the disc plane of a spiral galaxy,
valid over all but the innermost part of the disc plane,
\begin{equation}
u_0(r)=-1.4\frac{\rm km}{\rm s}\,
        e^{(r-R_\odot)/r_0} \left(\frac{r}{R_\odot}\right)^{-1}
        \left(\frac{\dot M}{1\,M_\odot\yr^{-1}}\right),\label{vr}
\label{inflow}
\end{equation}
where we put $r_0=5\kpc$.  With the density profile shown in
Fig.~\ref{Omeg}, the resulting mass accretion rate $\dot M\simeq2\pi
r\sigma u_0\approx 1 M_\odot\yr^{-1}$ is independent of radius within
20 per cent in $5<r<15$ kpc, and decreases at smaller $r$ (mimicking
a mass sink due to star formation in the galactic molecular ring).
In order to avoid large flow velocities in the inner part of the
disc, where observational data are poor and the situation uncertain,
we caused $u_0$ to go smoothly to zero between $r=r_u=1.5\kpc$ and
$r=0$.  Then if $u_0=-1\kms$ at the Solar position, we have
$u_0\approx -0.6\kms$ at $r=300\ppc$, which corresponds to $\dot
M\approx0.3\,M_\odot\yr^{-1}$ at that radius (for
$\sigma=300\,M_\odot\ppc^{-2}$), as given by Morris \& Serabyn
(1996).  We also explored the effect of reduction in the value of the
truncation radius $r_u$ (see Sect.~\ref{inflowh}).

The resulting radial velocity profile $u_0$ is shown in
Fig.~\ref{Omeg}. We also define a function
\[
f_u(z)=\cases{1\;,&$|z|\leq h\;,$\cr
\exp[-(|z|/h-1)^2]\;,&$|z|>h\;,$\cr}
\]
and put $u(r,z)=u_0(r)f_u(z)$, ensuring that the inflow is
essentially zero away from the disc.

We consider a range of models with various values of $u_0(R_\odot)$
in Eq.~(\ref{inflow}) by replacing the factor $1.4$ by $0.8U_0$ and
varying $U_0$; $u_0(R_\odot)\simeq-1\kms$ seems to be the most
plausible value.

\subsection{Introduction of dimensionless variables}
Equation~(\ref{mfd}) is nondimensionalized in terms of a length $R$
and a time $h^2/\eta_0$.  We measure angular velocity, radial
velocity and $\alpha$ in units of their typical values $\Omega_0$,
$U_0$ and $\alpha_0$ respectively, and $\vec{B}$ in units of
$B_0=(4\pi k\rho_0v_{\rm t}^2)^{1/2}$, an equipartition value with
respect to the turbulent kinetic energy.  The choice $\rho_0=5\times
10^{-24}\gcmcube$  corresponds to the maximum surface density given
by Sanders et al.\ (1984) outside the central region, which is about
$28\,M_\odot\ppc^{-2}$, with the effective scale height of the
neutral gas of $200$\,pc taken to be independent of $r$.  The gas
density $\rho$ was taken to be independent of $z$ in $|z|\leq h$; its
value in $|z|>h$ is irrelevant, as $\alpha=0$ there, and $\rho$ only
appears in the $\alpha$-quenching term.  We took
$\Omega_0=25\kms\kpc^{-1}$.

The poloidal and toroidal parts of Eq.~(\ref{mfd}) become respectively
\begin{eqnarray}
\frac{\partial A}{\partial t}&=&
        \left(\lambda R_\alpha\alpha\vec{B}+
        \lambda R_u u \widehat{\vec{r}}\times\vec{B}+
        \lambda^2\vec{u}_{\rm dia}\times\vec{B}\right)
        \cdot\widehat{\vec{\phi}}\nonumber\\
                &&\mbox{}\quad+\lambda^2\eta{\cal D}^2 A\;,
                                \label{compeqp}\\
\frac{\partial B}{\partial t}&=&
        \nabla\times(\lambda R_\alpha\alpha\vec{B}+\lambda R_u u
        \widehat{\vec{r}}\times\vec{B}+ R_\omega\Omega
        r\widehat{\vec{\phi}}\times\vec{B}\nonumber\\
          &&\mbox{}\quad+\lambda^2\vec{u}_{\rm dia}\times\vec{B})
                \cdot\widehat{\vec{\phi}}+\lambda^2\eta{\cal D}^2 B\;,
                                \label{compeqt}
\end{eqnarray}
where $A$ and $B$ are the dimensionless forms of $A_\phi$
and $B_\phi$, and
\begin{equation}
{\cal D}^2=\frac{\partial^2}{\partial
r^2}+\frac{1}{r}\frac{\partial}{\partial
r}-\frac{1}{r^2}+\frac{\partial^2}{\partial z^2}\;.
\end{equation}
The radial velocity in Eqs.~(\ref{compeqp}) and (\ref{compeqt}) is
measured in units of $U_0$, giving
\begin{equation}
u_0=-0.05(r/R)^{-1}\exp(r/r_0)\;.
\end{equation}

In these equations, and below, we refer to dimensionless quantities,
unless otherwise stated.  Thus we have the dimensionless parameters
\[
R_\alpha=\frac{\alpha_0 h}{\eta_0}\;,
\quad
R_\omega=\frac{\Omega_0 h^2}{\eta_0}\;,
\quad
R_u=\frac{U_0 h}{\eta_0}\;.
\]
We chose $R=15\kpc$, $Z=5\kpc$,
$h=500\ppc$ (the scale height of the ionized gas, distinct from that
of the neutral gas), $\eta_1=30$, and considered two values of the disc
magnetic diffusivity, $\eta_0=10^{26}$ and
$3\times10^{26}\cm^2\s^{-1}$.  The former (more conventional) value
of $\eta_0$ determines $R_\omega=18.75$, the latter $R_\omega=6.25$,
and both suggest that if $\alpha_0$ is of order $1\kms$, then values
of $R_\alpha$ of order unity are plausible; below we take
$R_\alpha=0.5$ and 2, respectively, both somewhat supercritical
values. With $\eta_0=10^{26}$ and $3\times10^{26}\cm^2\s^{-1}$,
values of the radial magnetic Reynolds number $R_u=3$ and $1$,
respectively, correspond to inward velocities of about $u_0=-1\kms$
at the solar radius of $10\kpc$.  With this value of $u_0(R_\odot)$,
the unit radial speed is $U_0\approx1.8\kms$.  If $v_{\rm t}=10\kms$
(at all radii), the unit of magnetic field is $B_0=8k^{1/2}\mkG$, and
the unit of magnetic energy is $\frac{1}{2} B^2_0R^2Z$.

\subsection{Boundary conditions}
At the boundaries $z=\pm \xi$, we used the boundary conditions
$\partial B/\partial z=0$, $B_r = -\partial A/\partial z = 0$.  If
the inner radius of the computational domain is at $r_{\rm min}=0$,
then regularity demands $A=B=0$ at $r=r_{\rm min}$.  For the many of
our computations we took, for numerical convenience, $r_{\rm
min}=0.02$, and applied boundary conditions on $A$ and $B$ that were
consistent with the analytical behaviour as $r\rightarrow 0$; most
importantly $\partial B_z/\partial r = 0$.  We also used $r_{\rm
min}=0$ for some computations -- comparison between solutions with
these values showed good agreement.  The boundary condition at $r=1$
is $B=0$, and the initial field is a small amplitude seed of mixed
parity.

We set $A=\frac{1}{2}{\cal{B}}$ at all times on the outermost radial
grid line $r=1$ with ${\cal B}=\mbox{const}$, and solve the
difference equations resulting from Eqs.~(\ref{compeqp}) and
(\ref{compeqt}) in the region $r_{\rm min}\leq r\leq 1$.  With
$B_z=r^{-1}\partial(rA)/\partial r$, this specifies a constant
vertical magnetic flux through the disc midplane $z=0$, given by
$F=2\pi\int_0^1B_zr\,dr\equiv\pi {\cal{B}}$ in dimensionless units.
This prescription means that at time $t=0$, when $A$ is negligible at
the inner grid lines, there is formally a discontinuity in $A$, and
so an infinity of $B_z$, near the boundary $r=1$.  However this is
immediately smoothed by diffusion, and the field is smooth after a
few time steps. Our interest is mainly with stationary magnetic
configurations (either steady or oscillatory) established at later
stages of evolution, especially as initial conditions are physically
uncertain.  We confirmed, by performing a trial integration with the
boundary condition that $B_z$ was uniform on the boundaries
$z=\pm\xi$, for a prescribed value of the flux, that our
computational procedure led to the unique solution.

Thus the interpretation of ${\cal{B}}$ is that it is the uniform
field in the $z$-direction that would have the same total flux
through the disc plane $z=0,\ r\leq 1$ as the model considered.

Equations (\ref{compeqp}) and (\ref{compeqt}) are discretized on a
grid with $N_r$ points uniformly distributed over the range $r_{\rm
min} \leq r \leq 1$ and $N_z$ points distributed over $-\xi \leq z
\leq \xi$. The spatial discretization is second order accurate in
space, and a second order Runge--Kutta method was used to advance the
solution in time. Our standard resolution was $N_r=51$, $N_z=201$,
but we used $N_r=401$ in the models with large central density,
$\sigma(0) = 300\,M_\odot\ppc^{-2}$.

\subsection{Diagnostic quantities}
Four key parameters are used to quantify the properties of our
stationary solutions, denoted by $E,\ P,\ P_{\rm i}$ and $f$. $E$ is
defined to be the total magnetic field energy in the computational
domain.

The global parity $P$ is defined as $P=(E_{\rm even}-E_{\rm odd})/E$,
where $E_{\rm even}$ and $E_{\rm odd}$ are the global energies of the
parts of the field possessing respectively quadrupole-like and
dipole-like symmetry with respect to $z=0$ (cf.\ Brandenburg et al.\
1992).  Thus a purely quadrupole-like field has $P=+1$, and $P=-1$
for a dipole-like field. We similarly define  $P_{\rm i}$, the field
parity in the region interior to $r_{\rm i}=0.02$ (300\,pc in
dimensional units). We monitor $P_{\rm i}$ to allow for the
possibility that dipolar magnetic field can dominate in the centre of
the disc (with a predominantly quadrupolar global parity), as
possibly observed in the Milky Way.  As the dominant toroidal
component of the magnetic field is strongly concentrated to the disc
region, values of $E$, $P$ and $P_{\rm i}$ calculated for the disc
region will be close to the values calculated for the complete
disc+halo system.

The vertical magnetic flux within radius $r$ is $F(r)=2\pi\int^r_0
B_zr\,dr = 2\pi A(r)r$, whereas the total magnetic flux through the
disc midplane is $F=F(1) = \pi {\cal{B}}$. Then $F_{\rm i}=F(0.02)$
is the flux within radius $r=0.02$ (arising necessarily from the
odd-parity part of the field), and the ratio $\varphi=F_{\rm i}/F$ is
the fourth descriptor of our results. Note that for a strictly
uniform field in the $z$-direction we have $A(r)\propto r$, and
$\varphi=r_{\rm i}^2=4\times 10^{-4}$.

To summarise, our results are presented in terms of the total
magnetic energy $E$, the global magnetic parity $P$, the parity $P_i$
of the field in the inner $300\ppc$ of the disc, and $\varphi$, the
relative vertical magnetic flux through the inner $300\ppc$.  The
average dimensionless vertical magnetic field within the inner
300\,pc can be obtained as $\overline{B}_{z,{\rm c}}
=2.5\times10^3{\cal B}\varphi $.

\section{The reference states}          \label{basic}
In this section we briefly describe two basic states of the system
considered -- firstly, dragging of an imposed vertical flux without
any dynamo action and, secondly, a disc dynamo without any imposed
flux or radial flow. These states provide references against which
the relative importance of dynamo action and advection in the
composite model can be assessed.  For the dynamo model described in
Sect.~\ref{puredyn} we take the lower of the values discussed for the
central gas density, $\sigma(0)=5\,M_\odot\ppc^{-2}$.  The
corresponding dynamo model for the high central density case,
$\sigma(0)=300\,M_\odot\ppc^{-2}$ is described in Sect.~\ref{dynh}.
The gas density does not enter the purely advective-diffusive
calculation discussed in Sect.~\ref{kin}.

\begin{figure}
\centerline{\includegraphics[width=7.5cm]{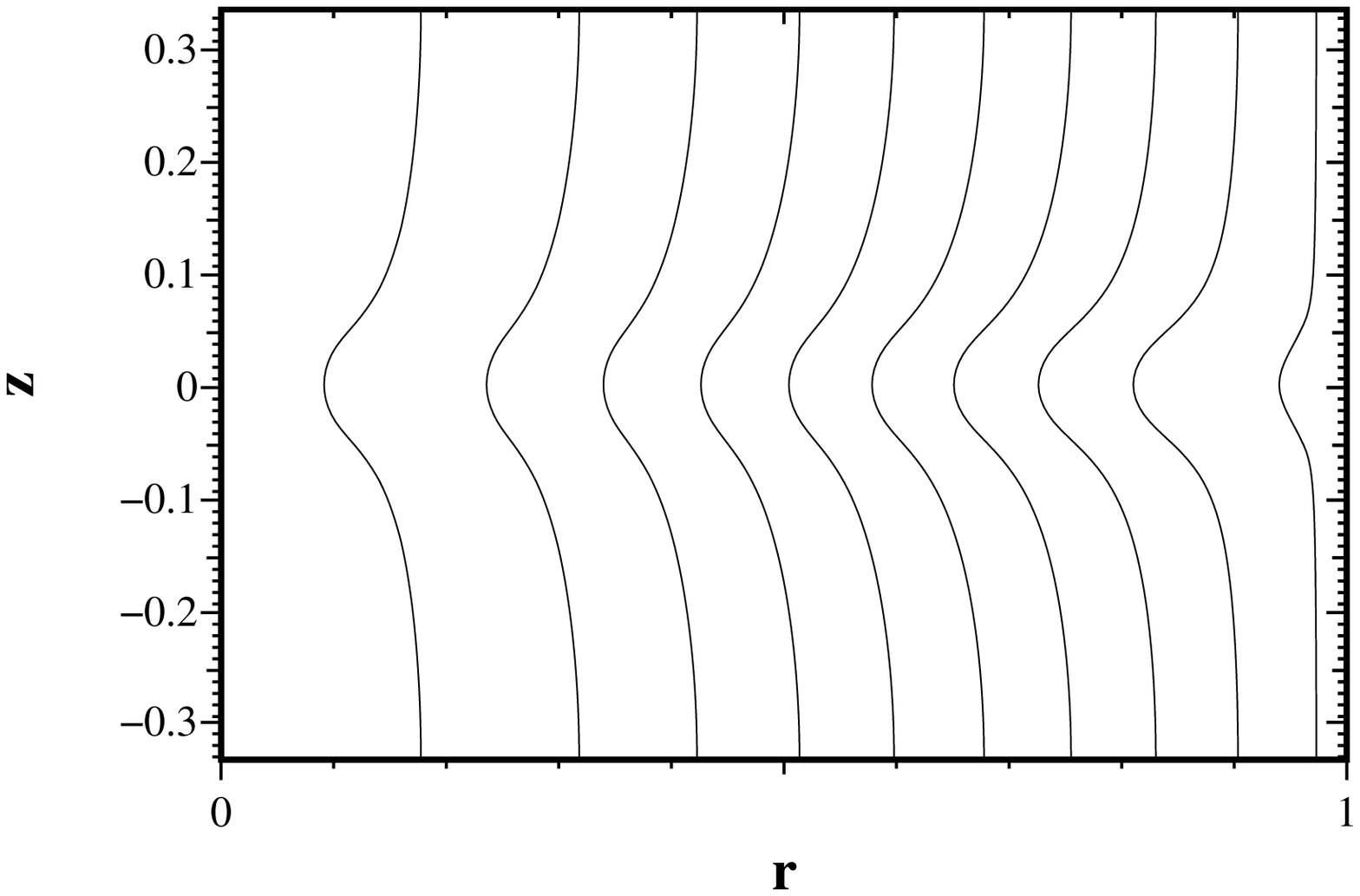}}
\centerline{\includegraphics[width=7.5cm]{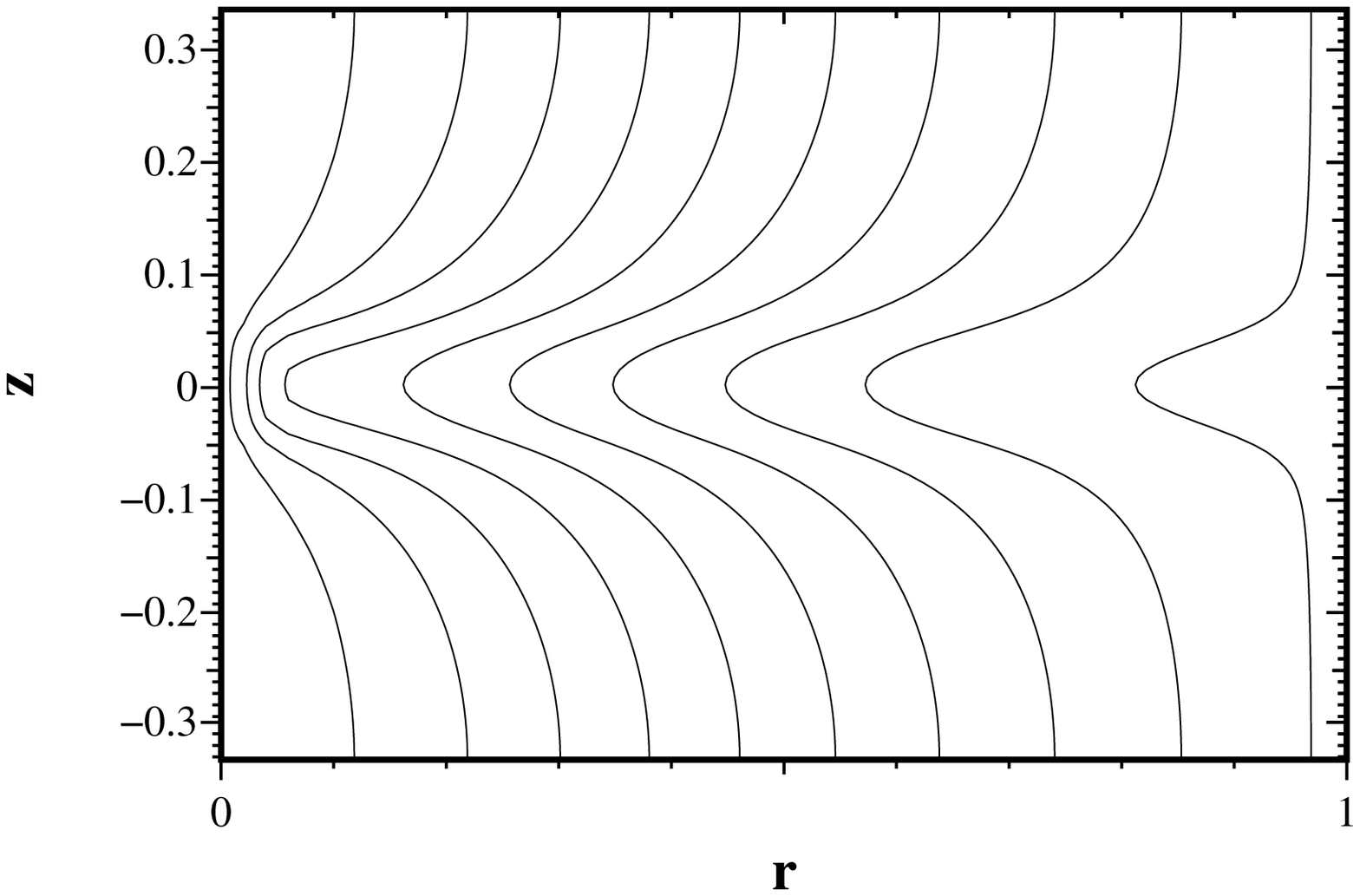}}
\caption[]{\label{fig_kin}
The poloidal field lines (equally spaced in poloidal magnetic flux
$rA$) in a steady state, for the kinematic advection calculation of
Sect.~\protect\ref{kin} with $R_\omega=18.75$,
$\eta_0=10^{26}\cm^2\s^{-1}$ $R_\alpha=0$, and (a) $R_u=1.5$, and (b)
$R_u=6$.
}
\end{figure}

\subsection{Magnetic flux captured by a dynamo inactive disc}
\label{kin}
We start by considering the advection of imposed vertical flux by a
radial inflow in  a differentially rotating disc in the absence of
dynamo action. Without the $\alpha$-effect in Eq.~(\ref{mfd}), the
equations (\ref{compeqp}) and (\ref{compeqt}) for $A$ and $B$
decouple and their solutions as $t\to\infty$ can differ from zero
only because ${{\cal{B}}} \neq0$.  So it is clear that, for given
values of $R_u$ and $R_\omega$, solutions will scale with the single
parameter ${\cal{B}}$, with magnetic energy $E_{\rm adv}\propto {\cal
{B}}^2$.  Equilibrium is determined solely by a balance of advection
and diffusion, and the toroidal field $B$ is parasitic on $A$, via
twisting of the poloidal field lines by the differential rotation and
vertical shear in $u$.  The field parity is odd, $P=-1$, as this is
the parity of the imposed external field.

We show in Fig.~\ref{fig_kin} the poloidal field lines for
$\eta_0=10^{26}\cm^2\s^{-1}$, $R_\omega=18.75$, and $R_u=1.5$ and 6.
The effects of increasing the magnetic Reynolds number of the radial
flow are clearly seen: field advection is markedly stronger when
$R_u$ is larger.

\begin{figure}
\centerline{\includegraphics[width=8.5cm]{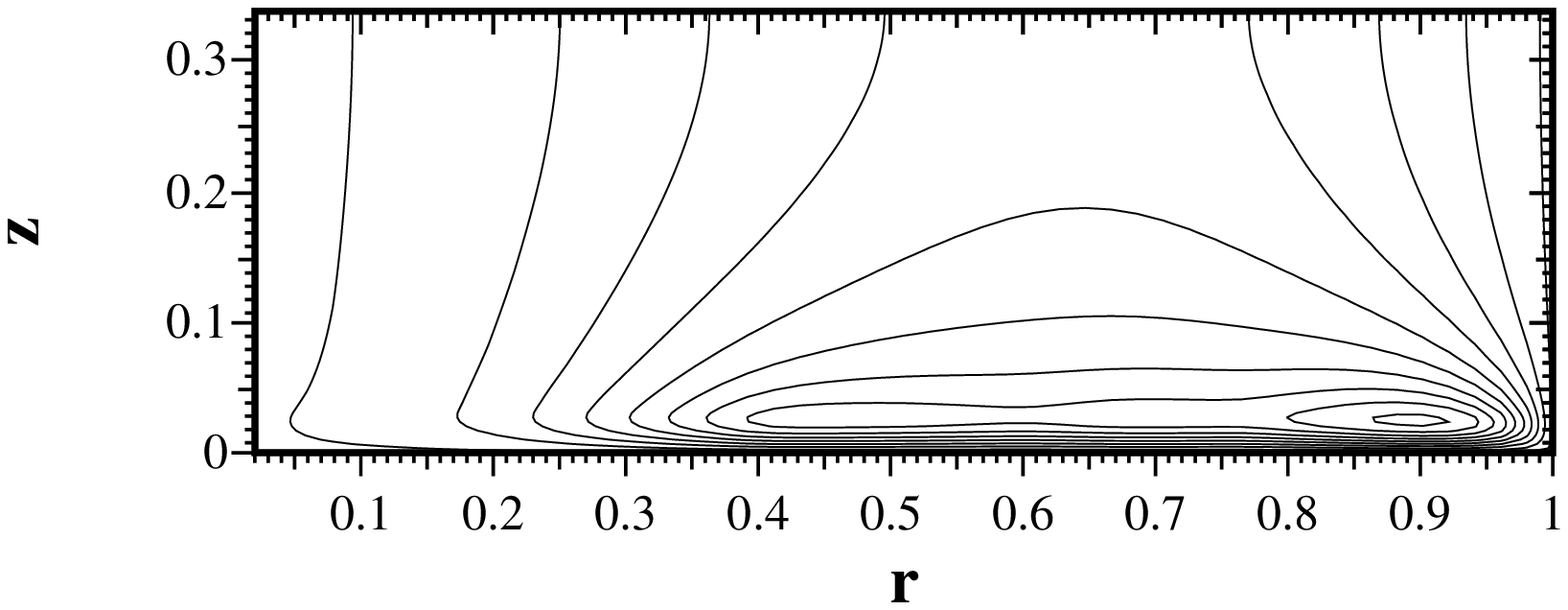}}
\centerline{\includegraphics[width=8.5cm]{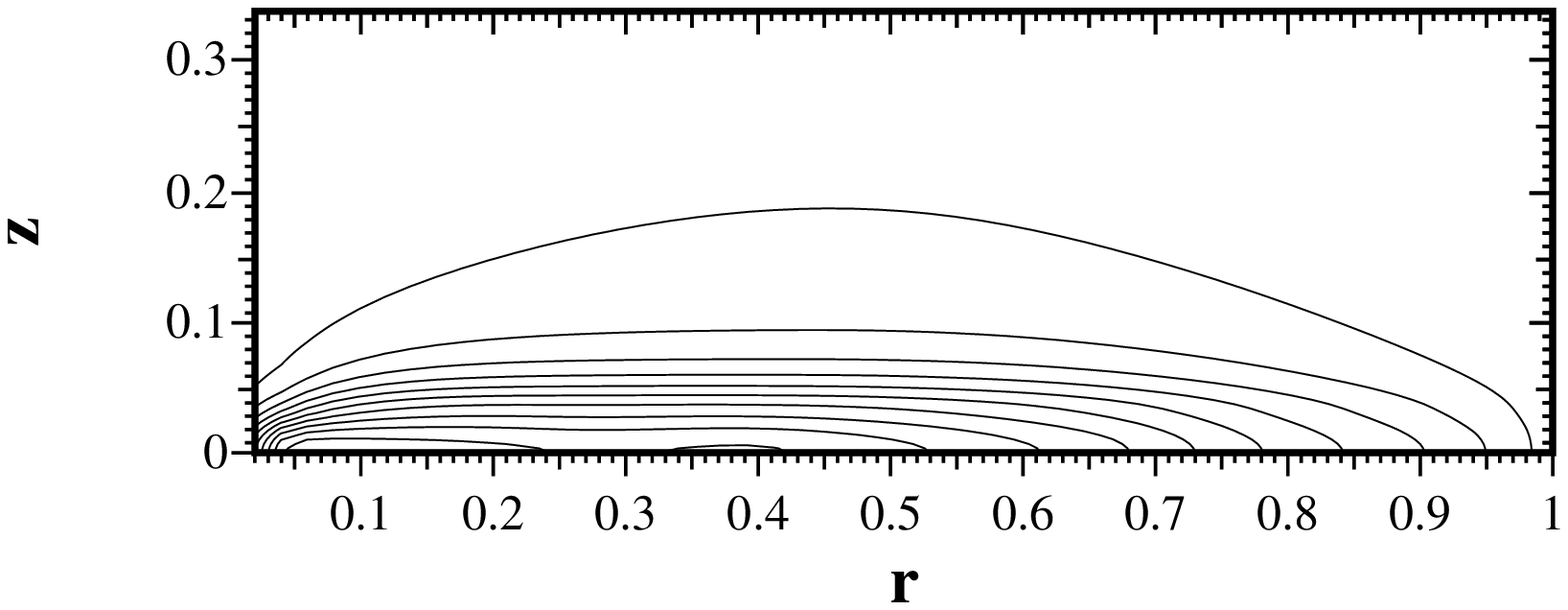}}
\caption[]{\label{dyn}
The poloidal field lines (a) and contours of toroidal field (b),
equally spaced in poloidal flux and toroidal field strength
respectively, for the pure dynamo calculation of
Sect.~\protect\ref{puredyn} with $R_u=0$, $R_\omega=18.75$,
$\eta_0=10^{26}\cm^2\s^{-1}$ and $R_\alpha=0.5$.  The field has an
even (quadrupolar, $P=+1$) structure with respect to the midplane
($z=0$), so only the upper half space is shown.
}
\end{figure}

\subsection{Disc dynamo without inflow}
\label{puredyn}
When $R_\alpha$ is non-zero, but $R_u=0$ and ${\cal{B}}=0$, we have a
standard disc dynamo. For reference, we show the field structure of a
nonlinear solution with our standard parameters $R_\alpha=0.5$,
$R_\omega=18.75$, $\eta_0=10^{26}$ cm$^2$ s$^{-1}$ in Fig.~\ref{dyn}.
The field parity is, as expected, quadrupolar ($P=+1$).  The toroidal
field is strongly concentrated in the disc, $|z|\leq 0.0333$, where
the $\alpha$-effect is non-zero, but the poloidal field pervades the
halo.  In this case we set ${\cal{B}}=0$ but, in the absence of an
inflow, the value of ${\cal{B}}$ is almost irrelevant.

\begin{figure}
\centerline{\includegraphics[width=8cm]{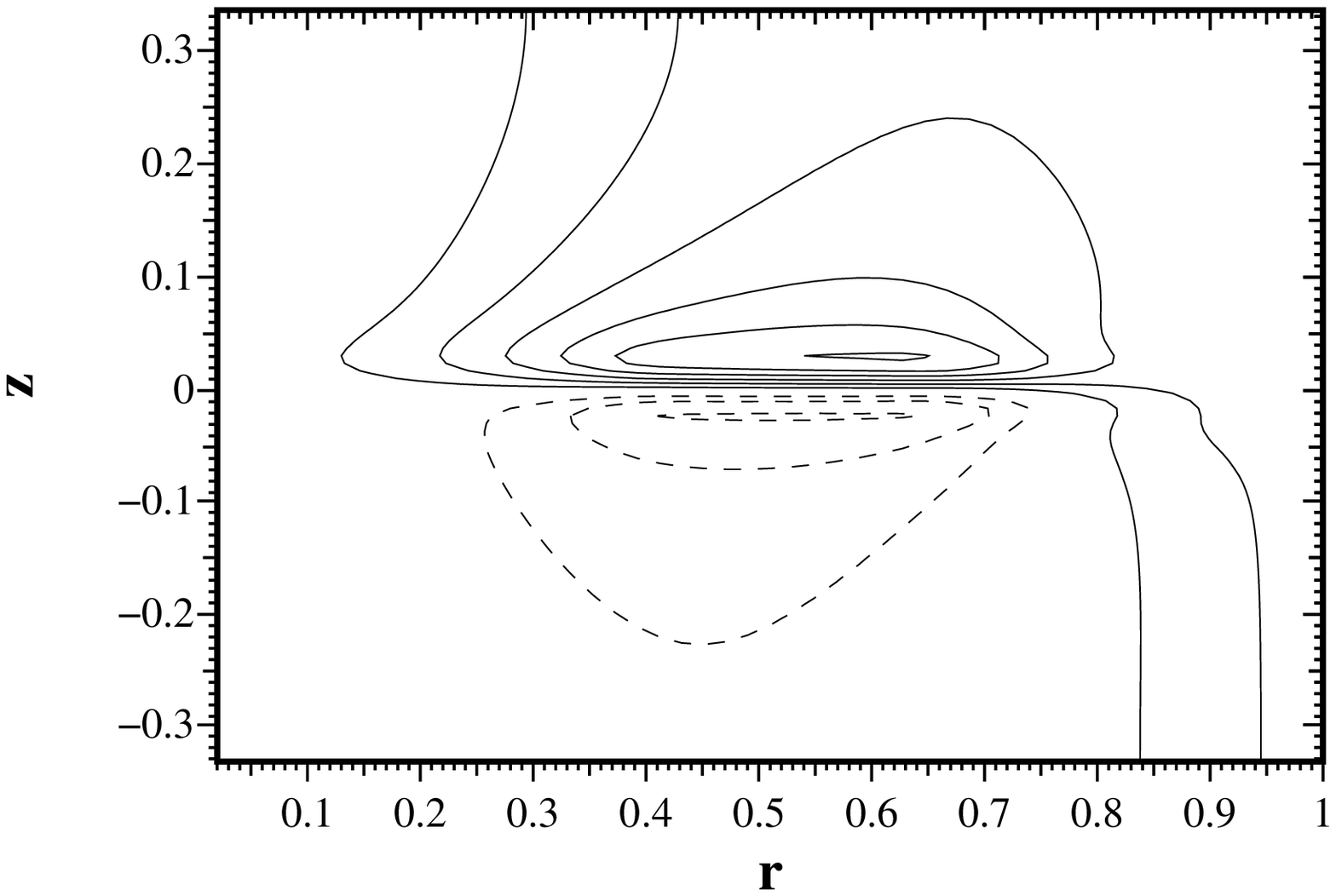}}
\centerline{\includegraphics[width=8cm]{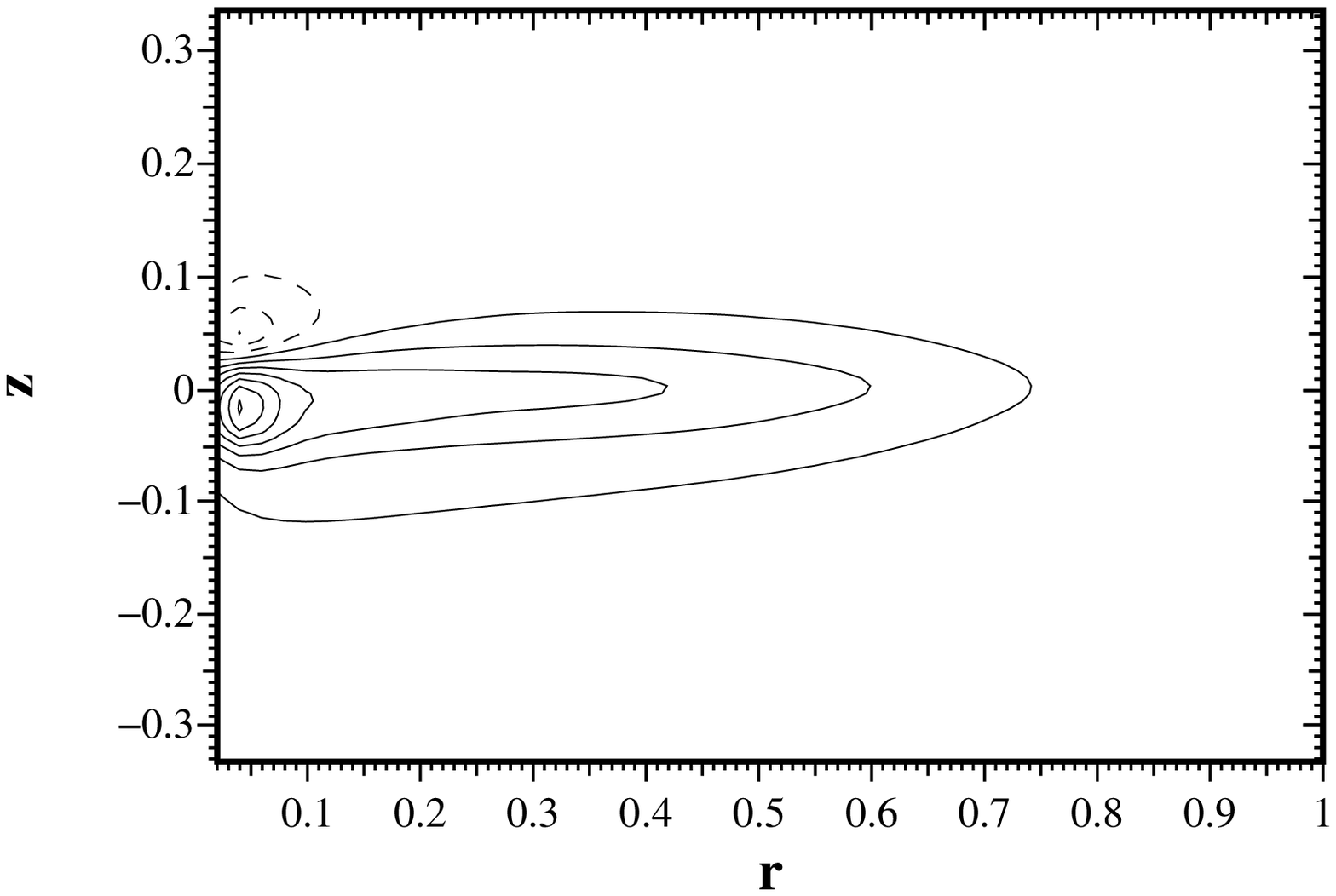}}
\caption[]{\label{mix2,01}
The poloidal field lines (a) and the contours of toroidal field (b),
both equally spaced, for the calculation of
Sect.~\protect\ref{standard} with $R_u=3$, $R_\omega=18.75$,
$\eta_0=10^{26}\cm^2\s^{-1}$, $R_\alpha=0.5$, ${\cal{B}}=10^{-3}$.
Continuous and broken contours refer to positive and negative values
respectively.
}
\end{figure}

\section{Dynamos with inflow; field advection and semi-dynamo action}
\label{inflowc}
In this section we discuss axisymmetric steady state magnetic fields
in a dynamo active disc where trapped vertical magnetic flux is
dragged in from large radii by an axially symmetric flow.  We explore
ranges of $R_u$ and ${\cal{B}}$, the two less certain parameters of
the model. The importance of the external magnetic field is also
sensitive to the magnetic diffusivity $\eta$ in the disc -- the
dragging is more efficient when the magnetic field is more strongly
coupled to the flow, i.e. when $\eta$ is smaller for a given inflow
speed. We consider two values of $\eta$, one being a standard value
for galactic discs, and the other a factor of three larger.

We first consider the `low' value for the central density,
$\sigma=5\,M_\odot\ppc^{-2}$, in order to allow the use of a lower
radial resolution.  In Sect.~\ref{standard} results from a basic
model, where we choose a single set of typical Galactic parameters,
are given.  Then we go on to explore how the results are affected by
changes in a few key parameters.  Results for our `standard' magnetic
diffusivity are presented in Sect.~\ref{standard} and those using an
enhanced diffusivity are in Sect.~\ref{altmodel}.

\subsection{Standard magnetic diffusivity: $\eta_0=10^{26}\cm^2\s^{-1}$}
\label{standard}
We begin by describing a typical nonlinear model with
$\eta_0=10^{26}\cm^2\s^{-1}$, choosing parameters ${\cal B}=10^{-3}$
(corresponding to a mean vertical field of $8\times10^{-9}k^{1/2}\G$
in imensional units), $R_u=3$, $R_\alpha=0.5$ and $R_\omega=18.75$.

We display in Fig.~\ref{mix2,01} the steady state magnetic field
structure found in this case.  There are clear deviations from pure
parity, either odd or even. The global parity parameter is $P=0.85$
indicating an almost quadrupolar parity, whereas in the inner 2\% by
radius $P_{\rm i}=0.02$. A small value of $|P_{\rm i}|$ means that
the magnetic field is relatively small either above or below the
disc, in this case in $z>0$ -- see Fig.~\ref{mix2,01}b.

\begin{table*}
\caption[]{\protect\label{results}
Summary of results for $\eta_0=10^{26}\cm^2\s^{-1}$, the central gas
surface density $\sigma(0)=5\,M_\odot\ppc^{-2}$ and $R_u=0,\ 1.5,\
3,$ and 6.  The two parity values given are respectively the global
parity $P$  and that  in the inner 2\% by radius, $P_{\rm i}$.  The
values of $E$ for ${\cal{B}}=0$ are those for the pure dynamo,
$E_{\rm dyn}$.  The magnetic energy obtained in the absence of dynamo
action, $E_{\rm adv}$, scales with ${\cal{B}}^2$, but is given
explicitly for convenience.  $\varphi=F_{\rm i}/F$ is the fraction of
the magnetic flux within radius $r=0.02$, and $\overline{B}_{z,{\rm
c}}$ is the dimensionless mean vertical field through $z=0$ within
radius $r=r_i=0.02$ (corresponding to the inner 300 pc in radius).
There are no entries for $\varphi$ when ${\cal{B}}=0$, as there is
zero flux through the midplane for a purely even parity field.  Where
a range of values are given, the solutions are oscillatory, and the
values are maxima and minima.  Note that the flux parameter $\cal{B}$
is the strength of the (dimensionless) uniform vertical field that
would have the same flux through the midplane as the corresponding
model. For reference, when $R_u=0$, ${\cal{B}}=0$, we find
$(P, P_i, E) = (+1, +1, 0.64)$.
}
\begin{flushleft}
\begin{tabular}{lcccccc}
\hline
${\cal{B}}$  &1 &$10^{-1}$ &$10^{-2}$ &$10^{-3}$ &$10^{-4}$ & 0\\
\hline
\multicolumn{7}{c}{$R_u=1.5$}\\[5pt]
$P$ & $-1$ & $-0.994$ & $0.17$ & $0.98$ & $0.9998$ & $+1$ \\
$P_{\rm i}$ & $-1$ & $-0.9999$ & $-0.82$ & $0.82$ & $0.997$ & $+1$ \\
$E$  & $1.13\times 10^3 $  & $1.68 $  & $0.80$  & $0.48$  & $0.48$  & $0.48$ \\
$E_{\rm adv}$  & $1.16\times 10^3$ & $1.16\times 10^1$& $1.16\times 10^{-1}$ &$1.16\times 10^{-3}$& $1.16\times 10^{-5}$& $0$\\
$\varphi$ &$4.5\times10^{-3}$ &$2.5\times10^{-3}$ &$5.7\times10^{-3}$ &$1.1\times10^{-2}$ & $1.2\times10^{-2}$ & -- \\
$\overline{B}_{z,{\rm c}}$&$11.3$ & $0.63$ & $0.14$ & $0.028$ & $0.0030$ & 0\\
\hline
\multicolumn{7}{c}{$R_u=3$}\\[5pt]
$P$ & $-1$ & $-1$ & $-0.55$ & $0.85$ & $0.998$ & +1\\
$P_{\rm i}$ & $-1$ & $-1$ & $-0.98$ & $0.02$ & $0.98$ & +1\\
$E$  & $7.56\times 10^3 $  & $48.4 $  & $1.08$  & $0.32$  & $0.30 $ & $0.32$ \\
$E_{\rm adv}$  & $7.60\times 10^3$ & $7.60\times 10^1$& $7.60\times 10^{-1}$ &$7.60\times 10^{-3}$& $7.60\times 10^{-5}$& $0$\\
$\varphi$ & $2.1\times10^{-2}$ & $2.1\times10^{-2}$ & $2.2\times10^{-2}$ & $4.5\times10^{-2}$ & $4.5\times10^{-2}$ & -- \\
$\overline{B}_{z,{\rm c}}$&$52.5$ & $5.3$ & $0.55 $ & $0.11 $ & $0.011$ & 0\\
\hline
\multicolumn{7}{c}{$R_u=6$}\\[5pt]
$P$ &$-1$ &$-1$ &$-1$ &$-0.35\ldots0.72$& $0.966\ldots0.997$ &+1\\
$P_{\rm i}$ &$-1$ &$-1$ &$-1$ &$-0.93\ldots0.78$&$0.45\ldots0.998$ &+1\\
$E$ &$4.08\times 10^4$ &$3.48\times10^2$ &$1.41\times10^{-1}$
&$(1.42\ldots14)\times10^{-3}$ &$(5.8\ldots128)\times10^{-4}$&$(5.6\ldots128)\times10^{-4}$\\
$E_{\rm adv}$  & $4.10\times 10^4$ & $4.10\times 10^2$& $4.10$ &$4.10\times 10^{-2}$& $4.10\times 10^{-4}$& $0$\\
$\varphi$ &$0.125$ &$0.117$ &$0.021$ &$0.0067\ldots 0.030$ &$0.0051\ldots 0.030$ &  --    \\
$\overline{B}_{z,{\rm c}}$&$312.0$ & $29.3$ & $0.53 $ & $0.017\ldots 0.075$ & $0.013 \ldots0.0075$ & 0\\
\hline \end{tabular} \end{flushleft}
\end{table*}

Our results are summarized in Table~\ref{results}.  The fields evolve
to steady configurations, except with the smaller ${\cal{B}}$ values
with $R_u=6$, when the dynamo is barely excited when ${\cal{B}}=0$
and oscillates for ${\cal{B}}=0,\ 10^{-3}$ and $10^{-4}$.

\begin{figure}
\centerline{\includegraphics[width=6.3cm]{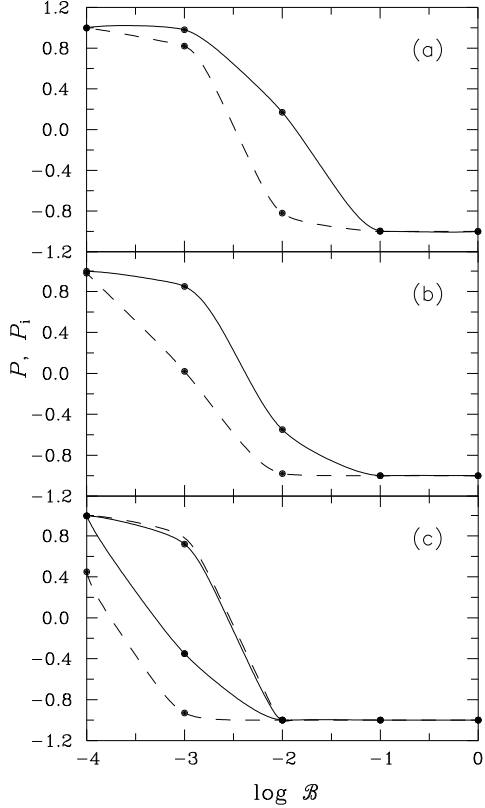}}
\caption[]{\label{Psummary}
The field parity in the whole disc, $P$ (solid), and in the inner
region of the disc $r<0.02$, $P_{\rm i}$ (broken), versus the
vertical flux parameter ${\cal{B}}$ for the calculation of
Sect.~\protect\ref{FPS} with $R_u=1.5$, $R_\omega=18.75$,
$\eta_0=10^{26}\cm^2\s^{-1}$, and $R_\alpha=0.5$. Panels (a), (b),
(c) are for the cases $R_\omega=1.5, 3, 6$ respectively. In (c), the
range of variation of the oscillatory solutions is indicated when
${\cal B}<10^{-2}$.  The field approaches purely odd parity, i.e.\
$P\to-1$, as ${\cal{B}}$ increases, but the field near the centre of
the disc has a systematically smaller value of $P$ (i.e., is more
dipole-like) than that in the disc as a whole, and needs a weaker
external field ${\cal{B}}\simeq10^{-2}$ to become approximately
dipolar.  }
\end{figure}

\begin{figure}
\centerline{\includegraphics[width=7cm]{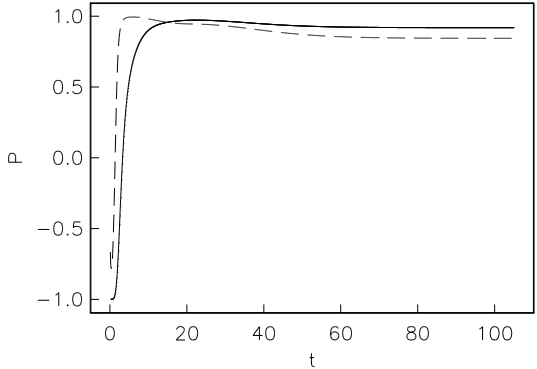}}
\caption[]{\label{parityab}
The parity of the poloidal field (solid) and toroidal field (broken)
in the whole computational domain as a function of time for the
calculation illustrated in Fig.~\protect\ref{mix2,01}.  (Time is
measured in units of $7.5\times10^8\yr$).}
\end{figure}

\subsubsection{Field parity and strength}\label{FPS}
Two trends are apparent. As illustrated in Fig.~\ref{Psummary}, for
given $R_u$, increasing ${\cal{B}}$ causes the parity to become
increasingly negative -- i.e.\ more dipole-like. However, this effect
is uniform neither in radius, nor between the poloidal and toroidal
field components. We show in Fig.~\ref{parityab} the evolution of the
parity of the poloidal and toroidal parts of the field for the
calculation illustrated in Fig.~\ref{mix2,01}. Of course, as the
energy in the toroidal part of the field is much greater than that in
the poloidal part, the parity of the toroidal field dominates the
global parity.  For a given ${\cal{B}}$ and $R_u$, the magnetic field
in the inner disc $r<0.02$ is more dipole-like than in the rest of
the disc as a stationary field configuration is approached (i.e.,
$P_{\rm i}$ is systematically smaller than $P$).  This also applies
to the lower values attained by $P$ and $P_{\rm i}$ in oscillatory
solutions.

Secondly, increasing $R_u$ decreases the dynamo efficiency, both in
the  linear and nonlinear regimes.  This is masked to some extent
when ${\cal{B}}\neq0$ by the increasingly efficient inward advection
of the vertical field.  When ${\cal{B}}=0$, we find that the total
dynamo-generated field energy decreases as $R_u$ increases:  the
trend is consistent with the findings of Moss et al.\  (2000) who,
however, mostly considered linear solutions.  We note that the
magnetic energy (mean taken when oscillatory) of these
two-dimensional solutions with ${\cal{B}}=0$  varies approximately
linearly with $R_u$ as
\begin{equation}
E_{\rm dyn}\approx E_0(1-R_u/6)\;,      \label{apprE}
\end{equation}
with $E_0=0.64$ for $\eta=10^{26}\cm\s^{-1},\ 0\leq R_u\leq6$ (see
Table~\ref{results}) and $E_0=0.90$ for
$\eta=3\times10^{26}\cm\s^{-1},\ 0\leq R_u\leq 5$ (see
Table~\ref{results1}; we also note that $E_{\rm dyn}=0.46$ for
$R_u=3$ consistent with this fit).

The result that advection inhibits dynamo action appears quite general;
Moss et al. (2000) produced an analytic argument, and our numerical results
arguably apply to an unfavourable case, as the advection tends to
sweep the field towards the central regions where the local growth rate is largest.

\begin{figure}[!thb]
\centerline{\includegraphics[width=6.5cm]{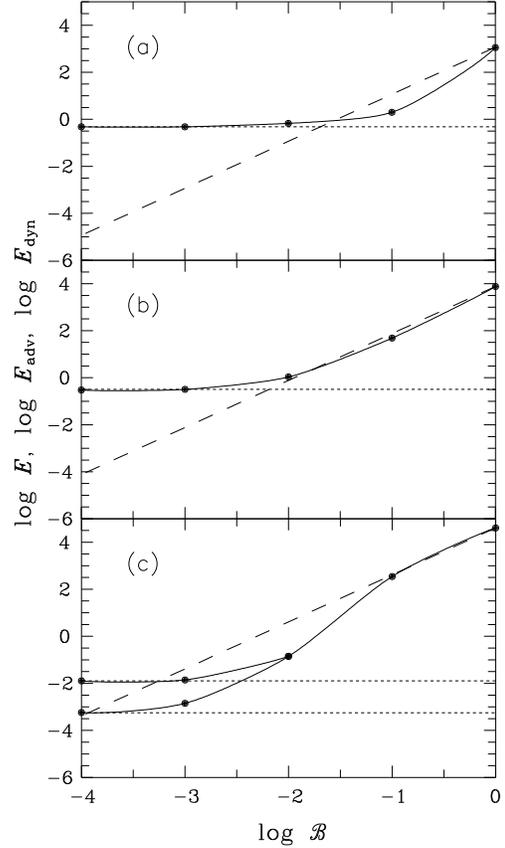}}
\caption[]{\label{Esummary}
The total field energy $E$, versus the flux parameter ${\cal B}$ for
the calculation of Sect.~\protect\ref{FPS}. $E$ is shown as  a solid
line for (a) $R_u=1.5$, (b) $R_u=3$ and (c) $R_u=6$. The other
parameters are $R_\omega=18.75$, $\eta_0=10^{26}\cm^2\s^{-1}$, and
$R_\alpha=0.5$. The solution for an advected field with no dynamo
action, i.e.\  $R_\alpha=0$, is shown by a broken line (cf.\
Sect.~\protect\ref{kin}), and that for a dynamo without inflow, i.e.\
$R_u=0$, is shown by a dotted line (cf.\
Sect.~\protect\ref{puredyn}). The field approaches that found in the
absence of dynamo action  as ${\cal{B}}$ increases, and that of a
pure dynamo (without inflow) as ${\cal{B}}$ decreases.  Some
solutions shown in panel (c) are oscillatory,  and then the solid and
dotted lines show the range of the field variation over a period.  }
\end{figure}

The dynamo and the inward advection of the vertical field combine in
a non-additive manner.  In Fig.~\ref{Esummary} (see also
Table~\ref{results}), we show the magnetic energy in a stationary
state, together with that resulting from pure advection
($R_\alpha=0$), and pure dynamo action (${\cal{B}}=0$).  The total
hybrid field energy $E$  consistently exceeds $E_{\rm adv}$ and is
close to $E_{\rm dyn}$ when ${\cal B}\leq10^{-2}$, whereas $E<E_{\rm
adv}$ for ${\cal{B}}=1$; we deduce that $E\to E_{\rm adv}$ as
${\cal{B}}\to\infty$.

\begin{figure}[!thb]
\centerline{\includegraphics[width=6.5cm]{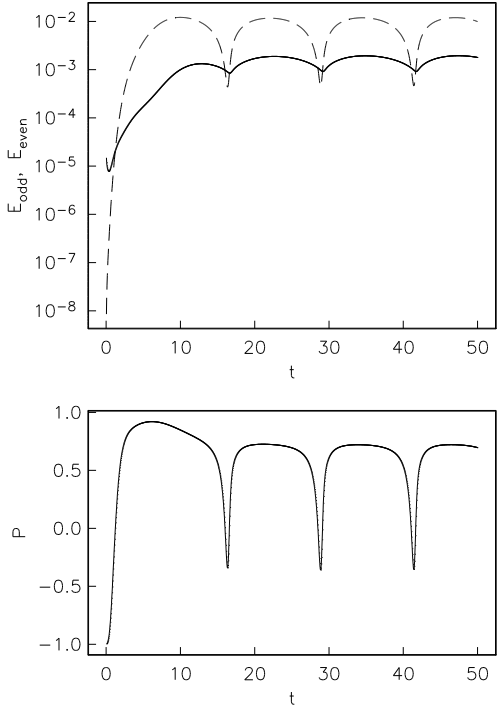}}
\caption[]{\label{Time}
Upper panel: Magnetic energies in the odd (solid) and even (broken) parts of
magnetic field in the whole computational domain for
$\eta_0=10^{26}\cm^2\s^{-1},\ {\cal{B}}=10^{-3}$ and $R_u=6$ (see
also Table~\protect\ref{results}). The solution is oscillatory but
its initial stage of growth is similar to that of non-oscillatory
solutions.
Lower panel: Global parity of magnetic field versus time.
The even part of the magnetic field oscillates at  a much
larger amplitude than the odd part. Time is measured
in units of $7.5\times10^8\yr$.
}
\end{figure}

The behaviour of the system when $R_u=6$ differs somewhat from that
for the smaller values of $R_u$. Now  an oscillatory dynamo is
excited when ${\cal{B}}=0$ (see Fig.~\ref{Time}).  The stationary
state has relatively small energy  and $E_{\rm adv}$ exceeds $E_{\rm
dyn}$ for ${\cal{B}} \ga 10^{-3}$.  Nevertheless, dynamo action plays
a significant role; in the range $10^{-3}\leq {\cal{B}} \la 1$ dynamo
action again reduces the magnetic energy of the system below that of
the purely advected field, so that $E<E_{\rm adv}$.  This case shows
some of the features of the model with
$\eta_0=3\times10^{26}\cm^2\s^{-1}$ and $R_u=10$ discussed in
Sect.~\ref{altmodel}.

The oscillatory behaviour of the solutions at large $R_u$ appears in
the linear solutions with ${\cal B}=0$ also. However the approximate
linear solution of Eqs.~(\ref{compeqp}) and \ref{compeqt}) developed
by Moss et al.\ (2000) in a one-dimensional approximation (retaining
only $z$-derivatives, see their Eq.~(A1)) remains non-oscillatory at
any $R_u$, and this is confirmed by their numerical solutions based
on the `no-$z$' approximation.  Thus the oscillations appear to arise
from the explicit fully two-dimensional structure of the model. The
period of these oscillations is of the order of the galactic age, so
they appear to be of little importance for galaxies, but their origin
in a broad dynamo theory context deserves further analysis.  We have
been unable to elucidate further either analytically or physically
the mechanism driving these oscillations.

\subsubsection{Field concentration towards the disc centre}
Another notable feature of the stationary states is that the fraction
of the vertical magnetic flux stored in the central part of the disc,
$r<0.02$, exhibits a non-trivial dependence on both $R_u$ and ${\cal
B}$.  Without dynamo action, $\varphi$, the fraction of the vertical
magnetic flux stored within $r=0.02$, increases with $R_u$.  These
values of $\varphi$ are identical to those shown in
Table~\ref{results} for ${\cal{B}}=1$, i.e.\  the solutions with
large ${\cal{B}}$ are close to those for pure advection.  Larger
values of $\varphi$ indicate stronger concentration of $B_z$ towards
the disc axis.  Note that $\varphi=4\times 10^{-4}$ for a uniform
vertical field.  For  dynamos without external vertical field (the
last column in Table~\ref{results}), $\varphi$ is not defined,
because the field has quadrupolar parity and hence has no flux
through the midplane $z=0$.

As illustrated in Table~\ref{results}, for $R_u=1.5$ (moderate
inflow), $\varphi$ unexpectedly grows as ${\cal{B}}$ decreases below
0.1. This growth is less pronounced for a stronger inflow, $R_u=3$.
The trend reverses as the inflow becomes very strong, with $\varphi$
decreasing as ${\cal{B}}$ decreases.  Thus, a larger trapped vertical
flux results in a weaker central concentration of the vertical
magnetic flux in the central part of a dynamo active disc provided
the inflow in not too strong.

For a fixed ${\cal{B}}$, the field is more  strongly concentrated to
the centre when the inflow is stronger -- a natural behaviour.  This
behaviour occurs for ${\cal{B}}\ga10^{-2}$ at all values of $R_u$
explored and for $R_u<6$ for    all values of ${\cal{B}}.$
Oscillatory solutions arising for $R_u=6$ behave differently having
smaller maximum values of $\varphi$ than steady solutions with the
same ${\cal{B}}$ but smaller $R_u$.

\subsubsection{Temporal variation}      \label{TV}
Since the boundary conditions at $r=R$ are inhomogeneous when
${\cal{B}}\neq0$, Eqs.~(\ref{compeqp}) and (\ref{compeqt}) do not
admit exponentially growing solutions even at the linear stage when
$B^2\ll4\pi k\rho v_{\rm t}^2$.  However, the quadrupolar part of the
magnetic field, maintained by the dynamo, exhibits initial
exponential growth at an early stage followed by saturation, as might
be expected for normal dynamo action. The dipolar dynamo modes decay
at the values of $R_\alpha$ and $R_\omega$ considered here, so the
dipole magnetic field within the disc only grows due to advection
(perhaps modified by semi-dynamo action), and this growth is
non-exponential. This is illustrated in Fig.~\ref{Time} where we show
the evolution with time of the energies in the even and odd parts of
the magnetic field for $R_u=6,\ {\cal{B}}=10^{-3}$.  The curve for
the even field deviates from a straight line in this log--linear plot
because of the contribution of the halo to the total energy (the
exponential growth is confined to the disc).  The solution is
oscillatory, and the quadrupolar part has a much larger amplitude of
oscillations than the dipolar; this is natural as the oscillations
are definitely driven by the dynamo. The initial growth before the
onset of the oscillations is fairly similar to that found in
non-oscillatory solutions.

The time for the steady state configurations to become established
will depend on the initial conditions. We have chosen what are
arguably the least favourable, with no flux initially penetrating the
disc. The time to reach a steady state is then of the order of
$10^{10}\yr$.  It is plausible that galaxies trap some poloidal flux
at all radii as they form.  Thus the timescale for our models to
reach a steady state should be regarded as an upper limit.  Moreover,
changes in the disc, being directly driven by the inflow, will occur
more rapidly if the inflow speed exceeds about $1\kms$, whereas the
halo adjusts on a slower, diffusive, timescale; therefore, a
stationary state can be established in the disc over a shorter time
scale.

\begin{table*}
\caption[]{\protect\label{results1}
Summary of results for calculations with $\eta_0=3\times
10^{26}\cm^2\s^{-1}$, the central gas surface density
$\sigma(0)=5\,M_\odot\ppc^{-2}$, and $R_u=0,\ 2,\ 5,$ and 10. The
format is similar to that of Table~\protect\ref{results}.
When $R_u=0$, ${\cal{B}}=0$, we obtain
$(P, P_i, E) = (+1, +1, 0.90)$.}
\begin{flushleft}
\begin{tabular}{lcccccc}
\hline
${\cal{B}}$  &1 &$10^{-1}$ &$10^{-2}$ &$10^{-3}$ &$10^{-4}$ & 0\\
\hline
\multicolumn{7}{c}{$R_u=2$}\\[5pt]
$P$ & $-1$ & $-0.7$ & $0.57$ & $0.99$ & $+1$ & $+1$\\
$P_{\rm i}$ & $-1$ & $-0.995$ & $-0.63$ & $0.91$ & $+1$ & $+1$\\
$E$  & $2.6 \times 10^2$  & $1.92$  & $0.78$  & $0.62$  & $0.60$ & $0.60$ \\
$E_{\rm adv}$  & $2.8\times 10^2$ & $2.8$ & $2.8\times 10^{-2}$& $2.8\times10^{-4}$& $2.8\times 10^{-6}$& $0$\\
$\varphi$ & $0.0080$ & $0.0057$ & $0.020$ & $ 0.024$ & $0.023$ &-- \\
$\overline{B}_{z,{\rm c}}$&$20.0$ & $1.43$ & $0.50 $ & $0.060 $ & $0.0058$ & 0\\
\hline
\multicolumn{7}{c}{$R_u=5$}\\[5pt]
$P$& $-1$ & $-1$ & $-0.18$ & $0.97$ & $+1$ & $+1$\\
$P_{\rm i}$& $-1$ & $-1$ & $-0.82$ & $0.80$ & $+1$ & $+1$\\
$E$  & $2.8 \times 10^3$  & $2.0$  & $0.62$  & $0.26$  & $0.26$ & $0.26$ \\
$E_{\rm adv}$  & $3.0\times 10^3$ & $3.0\times 10^1$& $3.0\times 10^{-1}$ &$3.0\times 10^{-3}$& $3.0\times 10^{-5}$& $0$\\
$\varphi$ & $0.064$ & $0.018$ & $0.077$ & $0.085 $ & $0.084 $ &-- \\
$\overline{B}_{z,{\rm c}}$&$160.$ & $4.50$ & $1.93 $ & $0.21 $ & $0.021$ & 0\\
\hline
\multicolumn{7}{c}{$R_u=10$}\\[5pt]
$P$ & $-1$ & $-1$ & $-1$ & $-1$ & $-1$ &    --\\
$P_{\rm i}$ & $-1$ & $-1$ & $-1$ & $-1$ & $-1$ &    --\\
$E$ &$7.0\times10^3$ &$3.2\times10^1$ &$3.0\times10^{-2}$ &$2.6\times10^{-4}$ &$2.6\times10^{-6}$ &decay\\
$E_{\rm adv}$ &$7.2\times10^3$ &$7.2\times10^1$ &$7.2\times10^{-2}$& $7.2\times10^{-4}$ &$7.2\times10^{-6}$ &$0$\\
$\varphi$ & $0.19$ & $0.13$ & $0.032$ & $0.022$ & $0.022$ &--   \\
$\overline{B}_{z,{\rm c}}$&$475.$ & $32.5$ & $0.80 $ & $0.055 $ & $0.0018$ & 0\\
\hline
\end{tabular}
\end{flushleft}
\end{table*}

\subsection{Enhanced magnetic diffusivity:
$\eta_0=3\times10^{26}\cm^2\s^{-1}$} \label{altmodel}
We then studied a model with an increased magnetic diffusivity in the
disc, $\eta_0=3\times10^{26}\cm^2\s^{-1}$ and thus also in the halo.
As in the previous section, we adopted $\sigma=5\,M_\odot\ppc^{-2}$
in $r<2\kpc$.  Now field amplification by differential rotation is
weaker because of enhanced diffusion, $R_\omega=6.25$, and so the
critical value of $R_\alpha$ is significantly increased; we took
$R_\alpha=2$, again a somewhat supercritical value for dynamo action.
With these parameter values, $R_u=1$ corresponds to an inflow speed
of about $1\kms$ near the Sun.  Our results are presented in
Table~\ref{results1}. Comparing these with those in
Table~\ref{results}, we see a broadly comparable pattern in the
results when $R_u\leq 5$, with advection significantly affecting the
dynamo field when ${\cal{B}} \gta 10^{-3}$.  Arguably, in this case
advection is somewhat less effective, consistent with what can be
expected for the larger magnetic diffusivity.  In particular, the
concentration of the vertical magnetic flux towards the disc axis,
$\varphi$, increases as ${\cal{B}}$ decreases for
$10^{-4}<{\cal{B}}<10^{-1}$, and decreases for smaller values of
${\cal{B}}$.

For $R_u=10$, the situation is rather different.  Now the dynamo is
not excited when ${\cal{B}}=0$.  Even so, dynamo action plays a role;
in the range $10^{-4}\leq B_{\rm ext} \leq 1$, we find that the
saturated magnetic energy is less than that of the advected field,
i.e.\ $E<E_{\rm adv}$, by a factor of approximately 3 for the smaller
values of ${\cal{B}}$. The field always has a dipolar parity, $P=-1$,
{\it not\/} typical of the dynamo configurations with moderate dynamo
numbers explored here.

When a quadrupolar field is not excited, the dipolar field energy in
the presence of an imposed flux and dynamo action is reduced from
that when the dynamo is turned off. This leads us to deduce that a
form of semi-dynamo, exciting an odd parity field, is operating (cf.\
Drobyshevski 1977); i.e.\ subcritical dynamo action that is dependent
on an external source of flux.  The fields so generated, although of
the same, odd, parity as the advected field, nevertheless interact
destructively with the advected field, reducing the total energy
below that of the purely advected field (i.e.\ that found with
$R_\alpha=0$).

\section{A Milky Way model with high central density} \label{SM}
Here we examine a model more specifically relevant to the Milky Way,
with $\sigma(0)=300\,M_\odot\ppc^{-2}$ and
$\eta_0=10^{26}\cm^2\s^{-1}$.  We take $R_u=3$, so the inflow
velocity near the Sun is about $1$ kms$^{-1}$.  The reference model
with no dynamo action is as described in Sect.~\ref{kin} -- in this
linear calculation the gas density has no role. We describe in
Sect.~\ref{dynh} a nonlinear dynamo model with no radial flow, before
giving our results for models with inflow and a range of ${\cal{B}}$
values.

\begin{figure}
\centerline{\includegraphics[width=8cm]{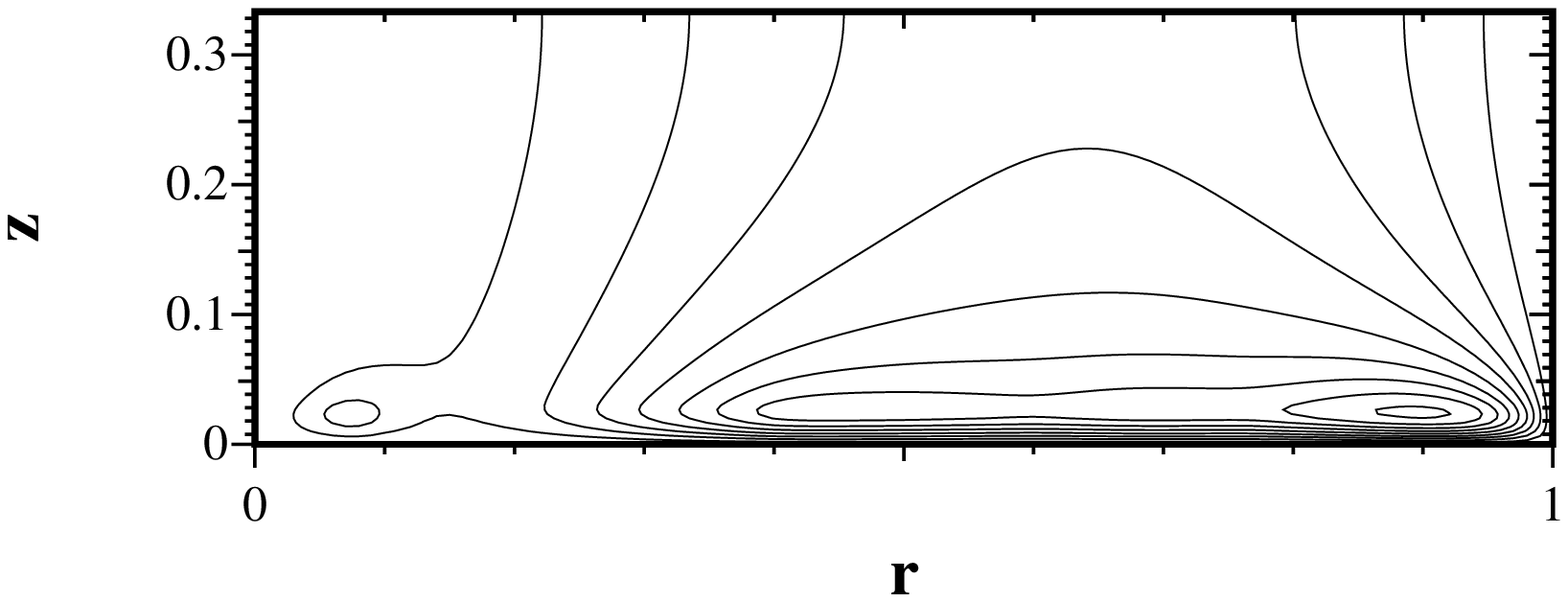}}
\centerline{\includegraphics[width=8cm]{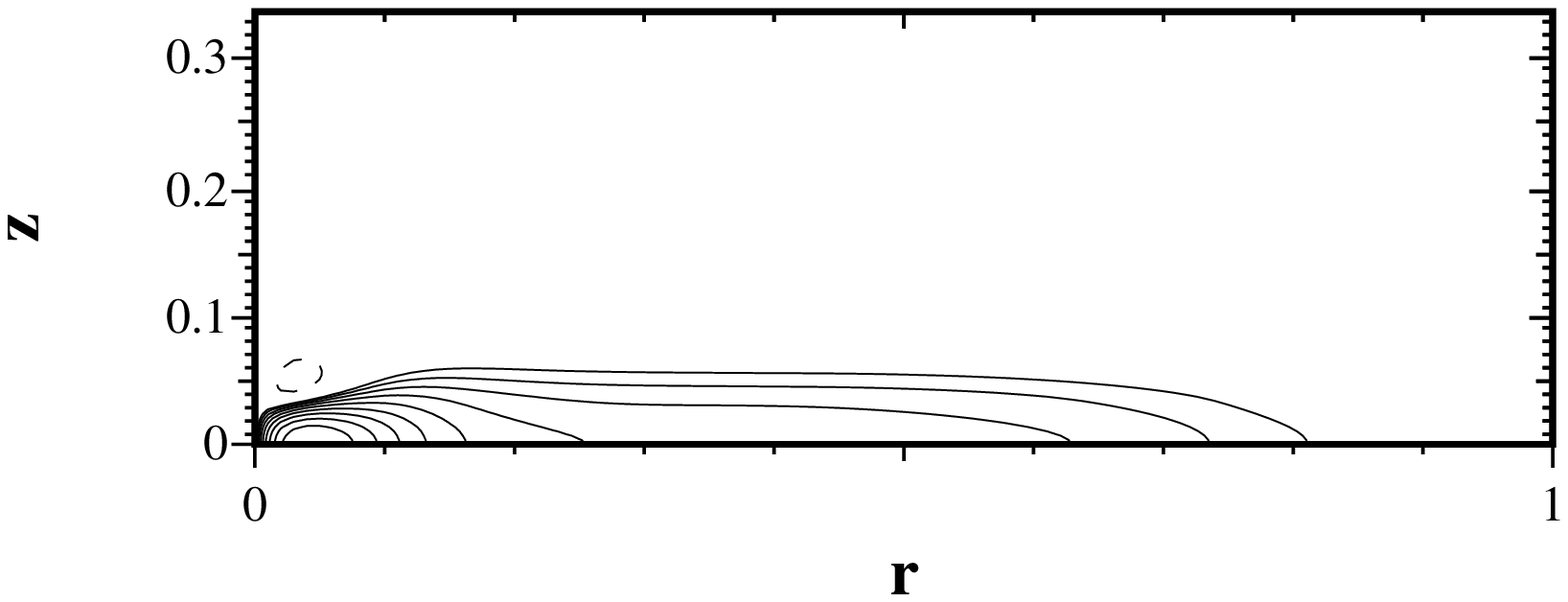}}
\caption[]{\label{dynhf}
(a) The poloidal field lines equally spaced in poloidal flux  and (b)
contours of toroidal field with contour levels starting at unity and
increasing by a factor of 1.5 in toroidal field strength between
contours, for the pure dynamo calculation of Sect.~\protect\ref{dynh}
with $R_u=0$, $R_\omega=18.75$, $\eta_0=10^{26}\cm^2\s^{-1}$,
$R_\alpha=0.5$ and $\sigma(0)=300\,M_\odot\ppc^{-2}$.  The field has
an even (quadrupolar, $P=+1$) structure with respect to the midplane
($z=0$), so only the upper half space is shown.}
\end{figure}

\subsection{The basic dynamo model}
\label{dynh}
In Fig.~\ref{dynhf} we show the poloidal field lines and toroidal
field contours for the basic dynamo model ($R_u=0$, ${\cal{B}}=0$).
Comparison with Fig.~\ref{dyn} shows that the toroidal field is much
more strongly concentrated to the galactic centre than in the models
of Sect.~\ref{inflowc} even though its magnitude at $r\simeq R_\odot$
remains approximately the same.  This is a natural consequence of the
strongly enhanced central density, and correspondingly increased
field strength $B_0$ at which the dynamo saturates.

\begin{table*}
\caption[]{\protect\label{results0}
Summary of results for $\eta_0= 10^{26}\cm^2\s^{-1}$, $R_u=3$,
and the central gas surface density
$\sigma(0)=300\,M_\odot\ppc^{-2}$.  The format is similar to that of
Table~\protect\ref{results}.
For $R_u=0$, ${\cal{B}}=0$ we obtain $(P, P_i, E) = (+1,+1, 0.84)$ when $R_\alpha=0.5$.
and $(+1, +1, 0.17)$ for $R_\alpha=0.25$.}
\begin{flushleft}
\begin{tabular}{lccccccccc}
\hline
${\cal{B}}$  &1 &$10^{-1}$ & $6\times 10^{-2}$ &$3\times 10^{-2}$&$2\times 10^{-2}$& $10^{-2}$ &$10^{-3}$ &$10^{-4}$ & 0\\
\hline
\multicolumn{10}{c}{$R_\alpha=0.5$}\\[5pt]
$P$ & $-1$ & $-1$ &$-1$ & $-0.86$ &$-0.69$ & $-0.35$ & $0.87$ & $0.990$ &$+1$\\
$P_{\rm i}$ & $-1$ & $-1$ &$-1$ & $-0.61$ & $-0.12$& $0.07$ & $0.82$ &$0.998$ & $+1$\\
$E$  & $7.66 \times 10^3$  & $48.0$ & $4.6$  & $0.80$ & $0.88$  & $0.80$  & $0.44$ &$0.42$  & $0.42$ \\
$E_{\rm adv}$  & $7.68\times 10^3$ & $7.68\times 10^1$ &$27.6$ & $6.91$ & $3.07$ &$0.768$& $7.68\times10^{-3}$& $7.68\times 10^{-5}$& $0$\\
$\varphi$ & $0.026$ & $0.021$ & $0.0089$ & $0.0032$ & $0.0051$  & $ 0.0080$ & $0.054$ & $0.067$ &-- \\
$\overline{B}_{z,{\rm c}}$&$65.0$& $5.2$ & $1.3$ & $0.24$ & $0.26$ & $0.20$ &$0.14$ &$0.017$& 0\\
\hline
\multicolumn{10}{c}{$R_\alpha=0.25$}\\[5pt]
$P$ & $-1$ &  & & $-1.0$ &$-0.996$ & $-0.56$ & $0.91$ & $0.999$ &$+1$\\
$P_{\rm i}$ & $-1$ &  & & $-1.0$ & $-0.992$& $0.12$ & $0.94$ &$0.999$ & $+1$\\
$E$  & $7.64 \times 10^3$  &  &  & $1.03$ & $0.38$  & $0.18$   &$0.11$&$0.11$  & $0.11$ \\
$E_{\rm adv}$  & $7.68\times 10^3$ & & & $6.91$ & $3.07$ &$0.768$& $7.68\times10^{-3}$& $7.68\times 10^{-5}$& $0$\\
$\varphi$ & $0.026$ &  &  & $0.0065$ & $0.0030$  & $ 0.0069$ & $0.025$ & $0.027$ &-- \\
$\overline{B}_{z,{\rm c}}$&$65.0$&  &  & $0.49$ & $0.15$ & $0.17$ &0.063 & $0.0067$& 0\\
\hline
\end{tabular}
\end{flushleft}
\end{table*}

\begin{figure}
\centerline{\includegraphics[width=8cm]{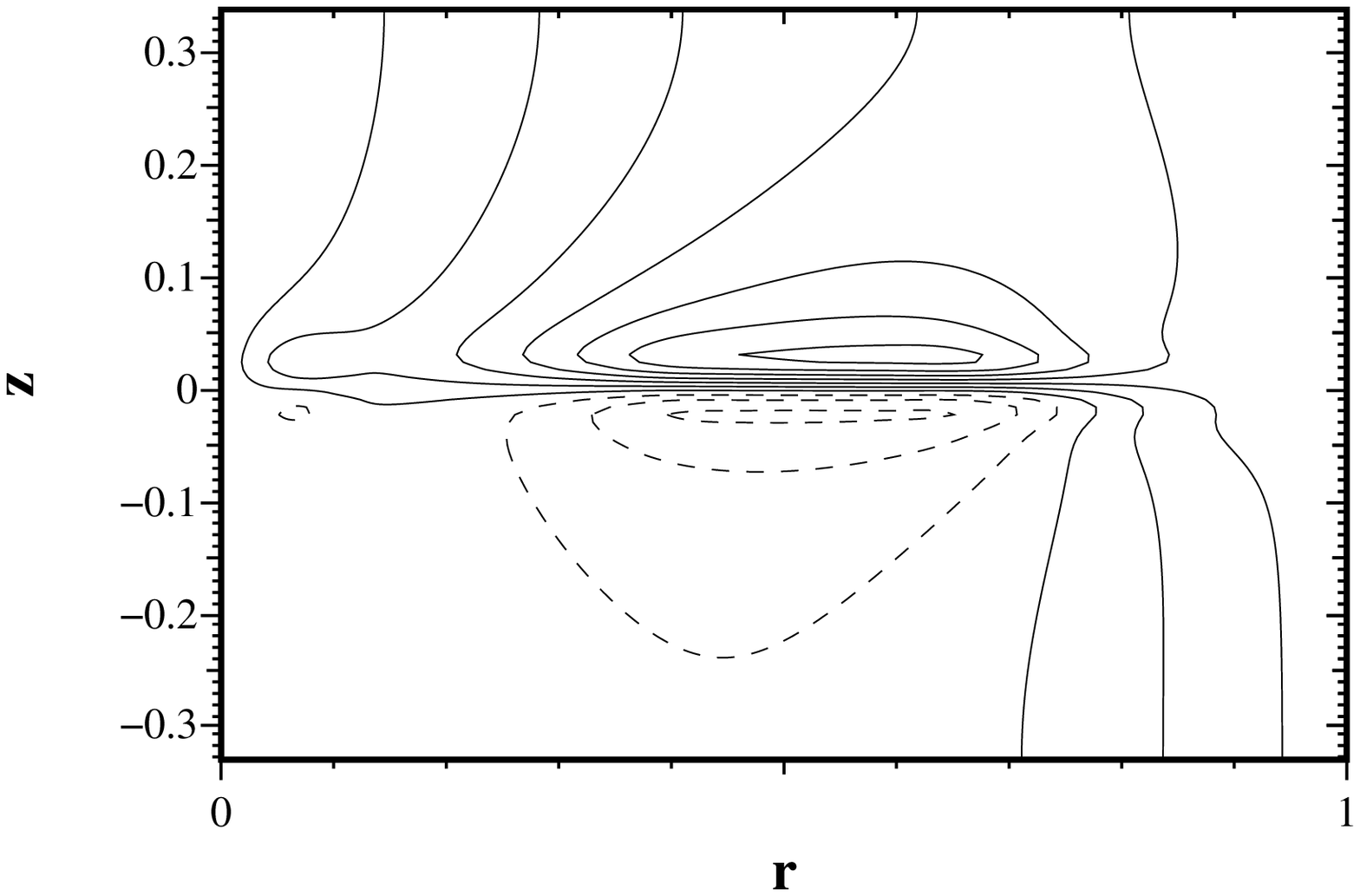}}
\centerline{\includegraphics[width=8cm]{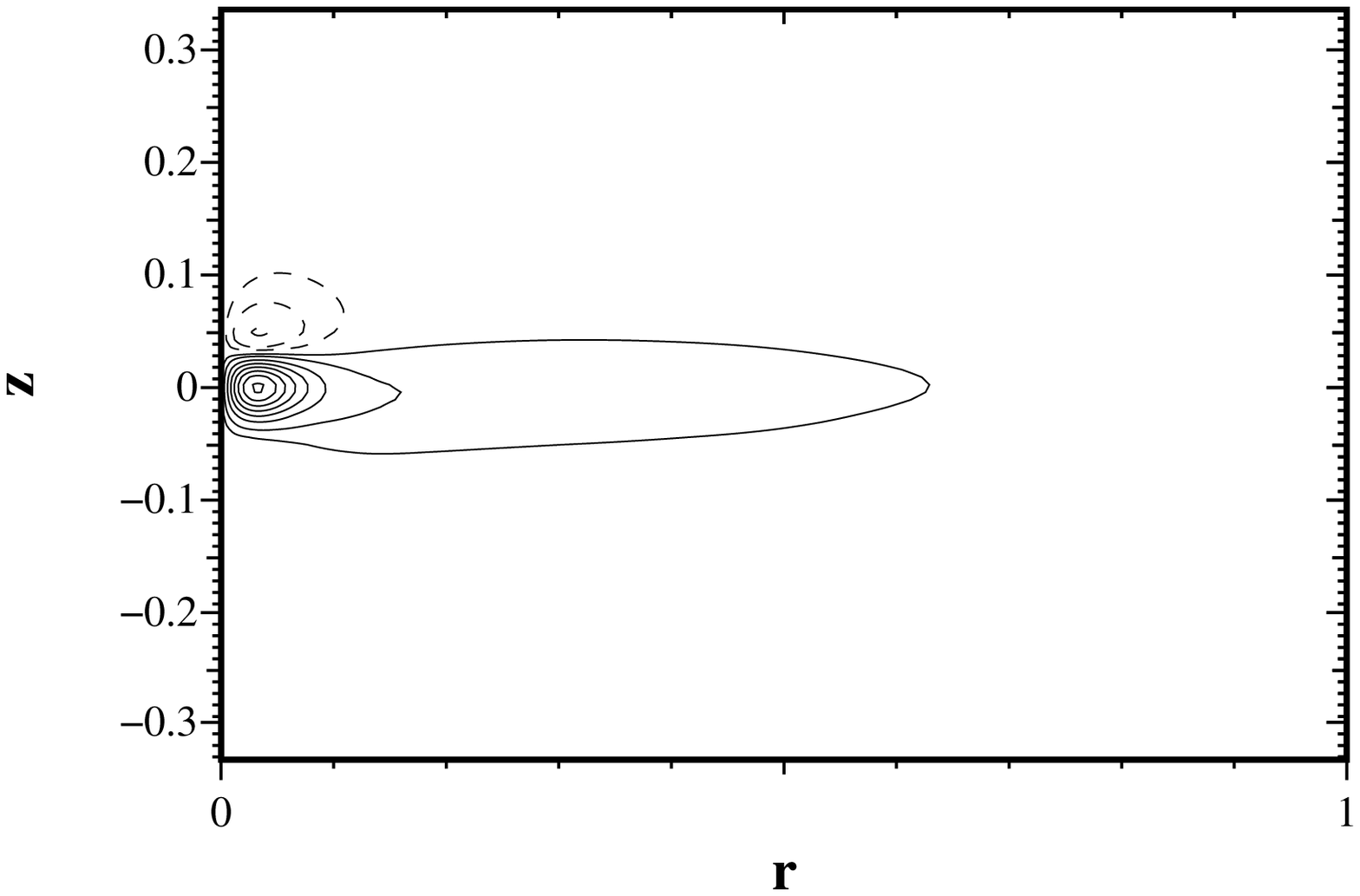}}
\caption[]{\label{figinflowh}
(a) The poloidal field lines equally spaced in poloidal flux  and (b)
contours of toroidal field with contour levels starting at unity and
increasing by a factor of 2.6 in toroidal field strength between
contours, for the calculation of Sect.~\protect\ref{inflowh} with
$R_\alpha=0.5$, $R_u=3$, $R_\omega=18.75$,
$\eta_0=10^{26}\cm^2\s^{-1}$, $R_\alpha=0.5$,
$\sigma(0)=300\,M_\odot\ppc^{-2}$ and  ${\cal{B}}=10^{-3}$.
Continuous and broken contours refer to positive and negative values
respectively.}
\end{figure}

\subsection{Dynamos with imposed flux through the midplane}
\label{inflowh}
We first discuss results with $R_\alpha=0.5$, the value used in
Sect.~\ref{standard}.  In Fig.~\ref{figinflowh} we present the field
structure (poloidal field lines and contours of toroidal field) for a
model with ${\cal{B}}=10^{-3}$.  Marked departures from pure symmetry
are visible; here $P=0.87$ and $P_i=0.82$, although these are not as
strong as for the corresponding model with small central density, see
Fig.~\ref{mix2,01}. The toroidal field is now much more strongly
concentrated to the galactic centre with
$B|_{r=0.02}/B|_{r=R_\odot}\simeq 20$.  Our results for $0\leq
{\cal{B}}\leq 1$ are given in Table~\ref{results0}.  We see a broad
similarity with the results of Sects.~\ref{standard} and
\ref{altmodel}.  The steady-state magnetic energy decreases with
increasing inflow rate, cf.\ Eq.~(\ref{apprE}). As ${\cal{B}}$
increases at fixed $R_u$, then $E_{\rm dyn}\leq E\leq E_{\rm adv}$
and both $P$ and $P_i$ decrease monotonically, so the field becomes
more dipole-like both globally and in the inner 300\,pc.  The only
significant difference to the results found above with the smaller
value of the central density is that $P_i<P$ for intermediate values
of ${\cal{B}}$, i.e.\  the central field is {\it less} dipole-like
than the global field.  Further (but related) we note that for these
intermediate values of ${\cal{B}}$, $\varphi$ is smaller than for the
corresponding model with $\sigma(0)=5\,M_\odot\ppc^{-2}$ (see
Table~\ref{results}).  We suspect this is because the stronger dynamo
generated field now present in the central regions is better able to
interact destructively with the advected field.

We also experimented with a reduction of the truncation radius $r_u$
for the velocity profile  to 0.05 (corresponding to $0.75\kpc$) from
our standard value $r_u=0.10$ (see Sect.~\ref{galmod}). This resulted
in an increase of $\varphi$ by a factor of about 2 when ${\cal
B}=10^{-3}$, and of about 2.5 when ${\cal B}=0.1$. We consider this
to give an estimate of the uncertainties in our model.

Given the importance of the destuctive interaction of dynamo and
advected fields, we also investigated a model with weaker dynamo
action, taking $R_\alpha=0.25$. These results are also presented in
Table~\ref{results0}.  The change of the diagnostic parameters as
${\cal B}$ increases follows a similar pattern as described above
when $R_\alpha=0.5$, except that the transition to a dipole-like
field structure occurs at smaller values of ${\cal B}$.  In
particular, we note that except when ${\cal B}=3\times 10^{-2}$ (when
already $P\approx -1$), the mean central field $\overline{B}_{z,c}$
is smaller than when $R_\alpha=0.5$.  We conclude that the dipolar
magnetic field at the disc centre is supported by dynamo action even
though the latter favours quadrupolar magnetic configurations. This
may be considered as another manifestation of semi-dynamo action, cf.
Sect.~\ref{altmodel}.

\section{Discussion and conclusions}
\label{disc}
\subsection{Implications for magnetic field in the Galactic centre}
An unexpected result of our simulations is that the relative
concentration of vertical flux in the central region, $\varphi$, does
{\it not\/} increase monotonically with the vertical
flux ${\cal{B}}$.  This effect is especially strong for moderate
inflow speeds.  For example, in Table~\ref{results0} with $R_u=3$,
$R_\alpha=0.5$, the fraction of the vertical flux stored in the inner
$300\ppc$ grows by a factor of 20 as the magnitude of the flux
parameter ${\cal{B}}$ {\it decreases} from  $3\times10^{-2}$ to
$10^{-4}$.  Increasing the central density enhances the central quadrupolar
field, but reduces the dipolar vertical field there.

\begin{figure}
\centerline{\includegraphics[width=7cm]{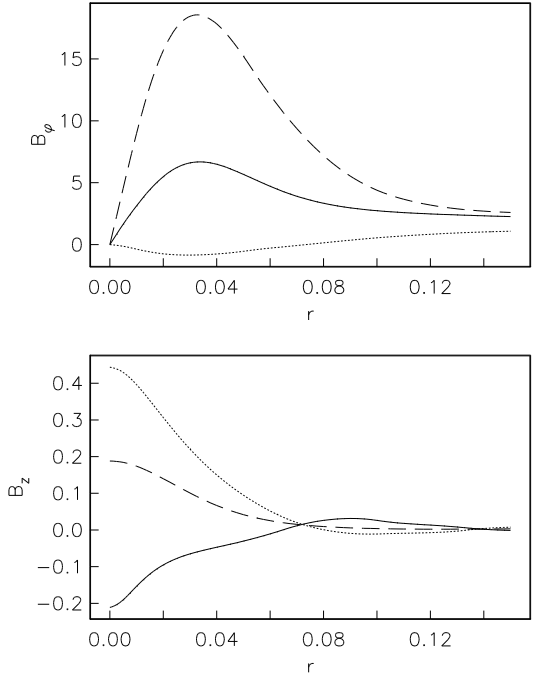}}
\caption[]{\label{radprofiles}
Radial profiles of the azimuthal (a) and vertical (b) magnetic field
components at $z=-500\ppc$ (solid), $z=0$ (dashed) and $z=500\ppc$
(dotted) in the inner $2\kpc$ in a model with $R_u=3$,
$R_\alpha=0.5$, $\eta=10^{26}\cm^2\s^{-1}$, the central gas surface
density $300\,M_\odot\ppc^{-2}$ and ${\cal{B}}=10^{-3}$ in
dimensionless units. Magnetic field is in units of $8k^{1/2}\mkG$,
and radius in units of $15\kpc$.}
\end{figure}

Our model, with a  high turbulent magnetic diffusion in the halo,
favours advection
(also the field is not anchored high above the
disc).  Nevertheless, our results indicate that the nonlinear
interaction with the dynamo limits the vertical magnetic field at the
centre to relatively low values even when both the advection and the
captured vertical field are rather strong.  In Tables~\ref{results},
\ref{results1} and \ref{results0} we give the values of
$\overline{B}_{z,{\rm c}} ={\cal B}\varphi/r_{\rm i}^2 =2500{\cal
B}\varphi$, the dimensionless mean vertical field within the inner
300 pc radius (i.e.\ $r\leq r_{\rm i}$), the overbar denoting a
horizontal average.  For the high central density model of
Sect.~\ref{SM} (Table~\ref{results0}), the mean central field
$\overline{B}_{z,{\rm c}}$ is approximately constant in an interval
$10^{-3}\la{\cal B}\la3\times10^{-2}$.  Quite generally,
$\overline{B}_{z,{\rm c}}\lta (10-20)k^{1/2}\mkG$ for ${\cal B} <
0.1$ and realistic values of $R_u$, and generally decreases with
${\cal{B}}$. (Note that this is a higher estimate than might be
deduced from Fig.~\ref{radprofiles}, where ${\cal B}=10^{-3}$.)

The estimated magnitude of $\overline{B}_{z,{\rm c}}$ may be further
increased by a factor of 2--3 because of uncertainty in the inflow
speed in the central region. However, any increased turbulent
velocity in $r<300\ppc$ (not allowed for in our models) would result
in larger values of $B_0$, thus both enhancing the quadrupole magnetic
field due to the dynamo and increasing the turbulent
diffusivity.  The latter would hamper advection of the
vertical magnetic field in the central regions, and could possibly
compensate any increase in $\overline{B}_{z,{\rm c}}$.  Another
possibility is that $\alpha_0$ may be underestimated at small radii
in our model if $\alpha_0\propto\Omega$; this would make the dynamo
stronger in the central regions, again favouring the generation of
quadrupolar magnetic fields there.  The resulting average vertical
field at $r<300\ppc$ would then remain not stronger than $0.01\mG$,
two orders of magnitude weaker than required if the magnetic field
illuminated in the filaments fills the volume.

\begin{figure}
\centerline{\includegraphics[width=8cm]{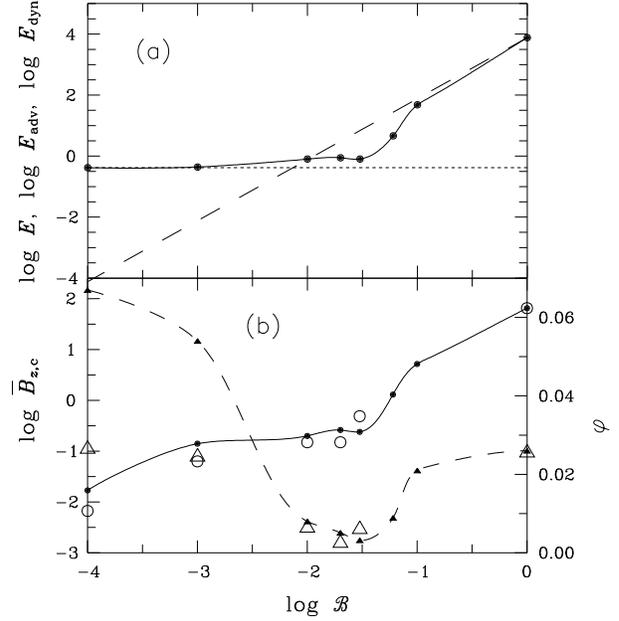}}
\caption[]{\label{MWsummary}
(a) The total magnetic energy $E$ versus the captured flux parameter
$\cal B$ in the model of Sect.~\protect\ref{SM} is shown by a solid
line for $R_\alpha=0.5$.  The advected field energy and dynamo energy
are shown dashed and dotted as in Fig.~\protect\ref{Esummary}.
(b)~The mean vertical magnetic field (solid) and the relative flux
$\varphi$ through the inner 300\,pc radius (broken) versus $\cal B$
for $R_\alpha=0.5$. The corresponding results for  weaker dynamo
action, with $R_\alpha=0.25$, are shown with open circles and triangles,
respectively.  Magnetic field is measured in units of
$8k^{1/2}\mkG$.}
\end{figure}

We show in Fig.~\ref{radprofiles} the radial variation of the
azimuthal and vertical magnetic field components in a model with a
radial velocity of about $1\kms$ near the Sun, where the degree of
concentration of $B_z$ to the centre, $\varphi$, is nearly the
strongest among the models studied. The profiles are shown for $z=0$
and $\pm500\ppc$ to provide further indication (in addition to the
parity parameter $P_{\rm i}$ introduced above) of the degree of
symmetry of the vertical profiles with respect to the midplane. Even
though the dipolar component [with $B_\phi(z)=-B_\phi(-z)$ and
$B_z(z)=B_z(-z)$] is significant (especially in the azimuthal field),
the dynamo-favoured quadrupolar component [with
$B_\phi(z)=B_\phi(-z)$ and $B_z(z)=-B_z(-z)$] is stronger.
Correspondingly, the parity in the inner 300\,pc is rather close to
unity (see Table~\ref{results0}).  Further details for these high
central density models are presented in Table~\ref{results0} and in
Fig.~\ref{MWsummary}.

The vertical magnetic field shown in Fig.~\ref{radprofiles} has a
maximum at $r=0$; this is a typical feature. However, the maximum
value is $B_z\simeq2\times10^{-6}\G$ at $r,z=0$ (taking $k=1$), which
is significantly below the level corresponding to equipartition with
kinetic energy, $B_0\simeq10^{-5}$--$10^{-4}\G$. The azimuthal field
vanishes at $r=0$ (because of the azimuthal symmetry) and reaches a
maximum at a distance of about 500\,pc from the centre; the maximum
value of about $1.5\times10^{-4}\G$ (again taking $k=1$) is slightly
above the equipartition level.

To limit the range of $\cal B$ of physical interest, we can apply
three constraints to our models.  On observational grounds we can
probably rule out solutions with $E\gg E_{\rm dyn}$, simply because
$E_{\rm dyn}$ is close to the kinetic energy of the turbulence and
corresponds well to the field strengths  observed in the outer parts
of the Milky Way and other galaxies.  Thus larger values of the
energy would imply strongly super-equipartition large scale (i.e.\
regular) fields throughout the galactic disc, for which there is no
evidence in normal spiral galaxies (e.g., Beck 2000).  This
constraint thus suggests that we should probably restrict our
attention to values ${\cal B}\la5\times 10^{-2}$.

Secondly, there is evidence to suggest that, in common with external
spiral galaxies, the global magnetic field of the Milky Way should be
of even (quadrupolar) parity (Frick et al.\ 2000 and references
therein). If conservatively we require  $P>0$,
then we need ${\cal B}\la10^{-2}$.

Thirdly, if we estimate the maximum large-scale magnetic field
strength in the protogalactic gas to be $\la3\times10^{-12}\G$
(Sciama 1994), then in contracting from 100\,kpc, an estimated radius
for the protogalaxy, to the present-day cylindrical radius of 20\,kpc
without loss of flux, we obtain a mean vertical field through the
disc of less than about $10^{-10}\G$, i.e.\ ${\cal B}\la10^{-5}$.
This is a negligibly weak field in the present context. For a galaxy
formed in a rich cluster of galaxies (though mostly populated by
ellipticals rather than spirals), the external field could be a
fraction of $10^{-6}\G$, corresponding to ${\cal B}\simeq
0.1$--$0.01$ in dimensionless units. This is clearly a very generous
upper limit for the Milky Way.

These considerations all suggest that we need only consider models
from Table~\ref{results0} with ${\cal B}\la0.05$ (of order
$10^{-7}$\,G), and possibly only those with ${\cal B}\la0.01$, or
even much less.  This implies that $\overline{B}_{z,{\rm c}}\lta
1.0$, that is the mean vertical field near the galactic centre is
less than about $10k^{1/2}\mkG$. Uncertainties associated with the
choice of the inflow velocity cut off parameter $r_u$ (see
Sect.~\ref{inflowh}) seem unlikely to change this estimate by more
than a factor of 2 or 3.  We also note that the same arguments
applied to the low central density models of Table~\ref{results} with
$R_u=3$ lead to similar estimates.

Dynamo action is not the only obstacle to the accumulation of dipolar
magnetic flux in the Galactic centre. Even the pure field advection
models with the lower value $\eta_0=10^{26}\cm^2\s^{-1}$ and
$R_\alpha=0$ as in Sect.~\ref{kin} (ignoring for a moment that these
models have strictly odd parity, $P=-1$ and so do not reproduce
magnetic field structure near the Sun), and ${\cal B}\la10^{-2}$
again yield $\overline{B}_{z,c}\la1$, and so the mean central field
is $\la10\mu$G.

The models explored here indicate that a vertical magnetic field of
an average milligauss strength cannot be accumulated in the centre of
the Galactic disc. The striking dipolar symmetry implied by the
straight shape of the magnetic filaments in the Galactic centre also
does not develop in our models.  Our model is not designed to produce
such fine filamentary structure. However, unless the filling factor
of the magnetic filaments is extremely small (i.e.\  the filaments
are not parts of a more uniform field illuminated only intermittently
-- see Morris \& Serabyn 1996 for a discussion), there is certainly
insufficient vertical flux present in our models.

We emphasize that we have only considered magnetic fields resulting
from Galactic-scale accretion and dynamo action. It cannot be
excluded that the central regions of the Galaxy represent an
autonomous system relatively weakly connected with the outer parts,
at least in terms of magnetic field. In that case, it would be
necessary to develop a separate, specific MHD model (allowing for
dynamo action) for this region. What we have shown, however, is that
the observed milligauss vertical magnetic fields within the inner
300\,pc would be difficult to explain by the accumulation of magnetic
flux from the outer parts if dynamo is active in the Galaxy, even if
the total vertical magnetic flux is conserved.  This
conclusion seems rather general and independent of the actual dynamo
mechanism.

Our conclusions are in variance with those of Chandran et al.\
(2000), principally because these authors assume the large-scale
magnetic field to be frozen into the interstellar gas. They argue
that the dissipation of magnetic energy in the hot gas, expected to
occur at the proton gyroradius scale, can take as long as
$10^{14}\yr$. However, the large-scale magnetic field is more
plausibly anchored in the warm phase (Beck et al.\ 1996) which has
larger filling factor and can be pervasive, whereas the hot gas is
buoyant and can leave the disc in a short time to be confined to
isolated regions in the disc.  Furthermore, the large-scale magnetic
field is affected mostly by {\it turbulent\/} magnetic diffusion
which is determined by the energy-range scale of turbulence about
$100\ppc$, and the velocity at this scale, rather than by the
dissipation scale of the turbulence.  Moreover, mean-field dynamo
action does {\it not\/} require genuine turbulence at all (i.e.\ any
details of turbulent energy cascade are unimportant) -- what is
needed is merely random, mirror-asymmetric, flow.  Just as
fundamentally, our results indicate that dynamo action in itself acts
to inhibit advection of field: this is the reverse side of inflow
inhibiting dynamo action in the absence of any external magnetic
field.  There is no reason to believe that this result is dependent
on the detailed nature of the dynamo (e.g.\ the $\alpha$-effect).

Chandran et al.\ (2000) similarly argue that gravity and tension of
the large-scale magnetic field will suppress the random nature of
interstellar gas motions thereby suppressing turbulent diffusion.  It
is, however, known that turbulent diffusion remains effective even in
stronger gravity fields, e.g., in the solar convective zone.
Magnetic fields undoubtedly suppress turbulent diffusion together
with the $\alpha$-effect (e.g., Krause \& R\"adler 1980), but these
effects plausibly occur simultaneously, so that dynamo action,
controlled by the dynamo number $D\propto\alpha/\eta^2$, can continue
(e.g.\ Moss in Mestel 1999, p.\  222; Brandenburg 2000.) Moreover, a
large-scale magnetic field can even enhance turbulent magnetic
diffusivity in some models (Brandenburg \& Subramanian 2000).
Efficient magnetic turbulent diffusion does not imply that gas will
be removed from the disc together with magnetic fields (Chandran et
al.\ 2000) as gas freely slips down the vertical sections of tangled
magnetic lines.

Altogether, we conclude that there are no compelling reasons to
consider the {\it large-scale\/} galactic magnetic field to be frozen
into the interstellar gas and the dynamo action to be suppressed.  We
reiterate that intense turbulent motions {\it are \/} observed in the
interstellar medium, and these motions cannot avoid tangling the
large-scale magnetic field.

\subsection{More general implications for galactic dynamo theory}
Our system also demonstrates several features of general interest in
dynamo theory.  In the main they have been described above, and are
merely recapitulated here.

For large enough inflow velocities, the dynamo is killed in the
absence of an imposed vertical flux, supporting the conclusions of
Moss et al.\  (2000).  As this situation is approached with
increasing $R_u$, the dynamo may become oscillatory at relatively low
energy; these  oscillations appear only in the fully 2D model studied
here, and not in the approximate dynamo solutions discussed by Moss
et al.\ (2000).  For small and moderate $R_u$, the steady-state
magnetic field scales as $(1-cR_u)^{1/2}$ with $c$ close to 1/6.

In the presence of a vertical flux (${\cal{B}} \neq 0$), semi-dynamo
action may occur; this is explicit when $R_u$ is large enough that
pure dynamo action is suppressed, and the steady-state magnetic
energy still differs significantly from that obtained with pure
advection, i.e.\ without any $\alpha$-effect.

In general the energy of the hybrid field (from dynamo action and
advection) is not the sum of the energies of the corresponding pure
dynamo and purely advected fields -- for larger ${\cal{B}}$ values,
$E < E_{\rm adv}$ even in cases where a dynamo is not excited when
${\cal{B}}=0$.  This is somewhat reminiscent of a result found by
Moss (1996) when investigating stellar dynamos with an imposed
background field.  We find that, as long as $E_{\rm adv}\gta E_{\rm
dyn}$, the (true) dynamo generated fields interact destructively with
the advected field, i.e.\ resulting in $E<E_{\rm adv}$.  In contrast,
the interaction is constructive when $E_{\rm adv}< E_{\rm dyn}$,
i.e.\ $E>E_{\rm adv}+E_{\rm dyn}$ in the stationary state.

Although it seems plausible that a large enough vertical flux and
inflow velocity could seriously inhibit dynamo action via
$\alpha$-quenching caused by the advected vertical field, we do not
see unambiguous evidence for this in our models. For large enough
$R_u$, the dynamo is in any case suppressed by essentially linear
effects (even when ${\cal{B}}=0$).  However, as $\cal{B}$ increases,
there is evidence ($E<E_{\rm adv}$) that the advected field promotes
some form of dynamo action, even in the extreme case $R_u=10$,
${\cal{B}}=1$ (Table~\ref{results1}).  It is clear that the effects
of nonlinearity in this dynamo system can take subtle forms.

We have here discussed only dynamo systems with $R_\alpha>0$, in
which the basic field generated is of even (quadrupolar) parity, as
believed to be appropriate to spiral galaxies. If in accretion discs
$R_\alpha$ is negative (Brandenburg et al.\ 1995, R\"udiger \& Pipin
2000), then a dipolar (odd) parity dynamo field is generated.  The
outcome of the interaction of such a dynamo with a trapped vertical flux of
the same odd parity is unclear. We intend to address this question
elsewhere.

\begin{acknowledgements}
We acknowledge support from  PPARC Grant PPA/G/S/1997/00284 and
NATO Linkage Grant PST.CLG 974737.
\end{acknowledgements}


\end{document}